\documentclass{aa}
\usepackage[varg]{txfonts}
\usepackage{hyperref}

\newcommand{\bc}{\begin{center}}
\newcommand{\ec}{\end{center}}
\newcommand{\be}{\begin{equation}}
\newcommand{\ee}{\end{equation}}
\newcommand{\bfig}{\begin{figure}}
\newcommand{\efig}{\end{figure}}
\newcommand{\m}{\mbox}

\newcommand{\oz}{O$_3$}
\newcommand{\om}{O$_2$}
\newcommand{\od}{O($^1$D)}
\newcommand{\ch}{CH$_4$}
\newcommand{\no}{N$_2$O}
\newcommand{\hox}{HO$_x$}
\newcommand{\nox}{NO$_x$}
\newcommand{\noo}{NO$_2$}
\newcommand{\hoo}{HO$_2$}

\begin{document}

\title{Is ozone a reliable proxy for molecular oxygen?} 
\subtitle{II. The impact of N$_2$O on the \om-\oz\ relationship for Earth-like atmospheres}

\author{Thea Kozakis \inst{1,2}
  \and Jo\~ao M.\ Mendon\c{c}a \inst{1,3,4} 
  \and Lars A.\ Buchhave \inst{1}
  \and Luisa M.\ Lara \inst{2}}
  
\institute{$^1$National Space Institute, Technical University of Denmark, Elektrovej, DK-2800 Kgs. Lyngby, Denmark\\
$^2$Instituto de Astrofísica de Andalucía - CSIC, Glorieta de la Astronomía s/n, 18008 Granada, Spain \\
$^3$School of Physics and Astronomy, University of Southampton, Highfield, Southampton SO17 1BJ, UK \\
$^4$ School of Ocean and Earth Science, University of Southampton, Southampton, SO14 3ZH, UK}

\date{}							

\abstract{

Molecular oxygen (\om) will be an important molecule in the search for biosignatures in terrestrial planetary atmospheres in the coming decades. In particular, \om\ combined with a reducing gas (e.g., methane) is considered strong evidence for disequilibrium caused by surface life. However, there are circumstances where it would be very difficult or impossible to detect \om, in which case it has been suggested that ozone (\oz), the photochemical product of \om, could be used instead. Unfortunately, the \om-\oz\ relationship is highly nonlinear and dependent on the host star, as shown in detail in the first paper of this series. This paper further explores the \om-\oz\ relationship around G0V-M5V host stars, using climate and photochemistry modeling to simulate atmospheres while varying abundances of \om\ and nitrous oxide (\no). Nitrous oxide is of particular importance to the \om-\oz\ relationship not only because it is produced biologically, but because it is the primary source of nitrogen oxides (\nox), which fuel the \nox\ catalytic cycle, which destroys \oz\ and the smog mechanism that produces \oz. In our models we varied the \om\ mixing ratio from 0.01-150\% of the present atmospheric level (PAL) and \no\ abundances of 10\% and 1000\% PAL. We find that varying \no\ impacts the \om-\oz\ relationship differently depending strongly on both the host star and the amount of atmospheric \om. Planets orbiting hotter hosts with strong UV fluxes efficiently convert \no\ into \nox, often depleting a significant amount of \oz\ via faster \nox\ catalytic cycles. However, for cooler hosts and low \om\ levels we find that increasing \no\ can lead to an increase in overall \oz\ due to the smog mechanism producing \oz\ in the lower atmosphere. Variations in \oz\ result in significant changes in the amount of harmful UV reaching the surfaces of the model planets as well as the strength of the 9.6~$\mu$m \oz\ emission spectral feature, demonstrating potential impacts on habitability and future observations.
}


\titlerunning{\oz\ as a Proxy for Molecular Oxygen II}
\authorrunning{Kozakis et al.}

\maketitle

\nolinenumbers

\section{Introduction}
Numerous studies have suggested the use of ozone (\oz) as a proxy for molecular oxygen (\om) in recent decades (e.g., \citealt{lege93,desm02,segu03,lege11,mead18}). This is largely because \om\ when combined with a reducing species such as methane (\ch), is considered a strong disequilibrium biosignature. Observing both in a terrestrial planetary atmosphere indicates that strong replenishing fluxes of \om\ and \ch\ must be present. Currently, life is the only known mechanism capable of providing these replenishing fluxes (e.g., \citealt{love65,lede65,lipp67,mead17}). Ozone comes into the picture because there are scenarios in which \om\ would be very difficult or impossible to detect in a planetary atmosphere when \oz\ is readily detectable. For example, the mid-IR is an excellent wavelength range for atmospheric characterization, due to the strong spectral features of potential biosignatures (e.g., \citealt{quan21,ange24}) along with the lessened impact of clouds in planetary emission spectra (e.g., \citealt{kitz11}). However, there are no strong \om\ features in the mid-IR -- only a collisionally induced absorption feature, which is sensitive only to large amounts of abiotically produced \om, not to the smaller amounts indicative of life \citep{fauc20}. Another example considers a planetary atmosphere resembling the low \om\ environment of early Earth, rather than the oxygen-rich atmosphere of modern Earth (\om\ comprising 21\% of the atmosphere). Observing a planet with \om\ abundances expected from the Proterozoic era on Earth (2.4-0.54 Gyr ago) \om\ would be difficult to detect, while \oz\ may be detected at trace amounts (e.g., \citealt{kast85,lege93,desm02,segu03,lege11}). For these reasons \oz\ has been seen as a good alternative to \om.

However, although \oz\ is the photochemical product of \om, it has also been known for decades that the \om-\oz\ relationship is highly nonlinear, due to both the pressure and temperature dependency of \oz\ formation and the requirement of UV for \om\ photolysis. To study this further, in the previous paper of this series, \cite{koza22}, we performed atmospheric modeling of Earth-like planets for a range of \om\ abundances and a variety of host stars. Here, we use the term “Earth-like” to refer to a terrestrial planet that is the same size and density as Earth, possesses a similar atmospheric composition, and orbits at a distance where it receives the same total flux from its host star as modern Earth. In our first study, we varied the atmospheric abundance of \om\ from 0.01-150\% PAL (present atmospheric level) around G0V-M5V host stars. We found that the \om-\oz\ relationship was extremely influenced by the host star, with planets around hotter stars (G0V-K2V) following different trends to those around cooler stars (K5V-M5V). Planets around hotter stars reached their peak \oz\ abundance at \om\ levels of just 25-55\% PAL, with the Earth-Sun system having a similar amount of \oz\ at both 10\% and 100\% PAL \om. This was due to the pressure dependency of \oz\ formation, an effect first discussed in \cite{ratn72}, although our previous study was the first to replicate it using modern atmospheric models and host stars other than the Sun. Cooler stars, on the other hand, were shown to host planets on which \oz\ decreased along with \om. Additionally, we found that the amount of UV reaching the planetary surface varied nonlinearly with both incoming UV and \oz, and that \oz\ features in simulated emission spectra depended more strongly on atmospheric temperature profiles than on the actual amount of \oz\ (or \om) in the atmosphere. Already from that study, we determined that using \oz\ as a proxy for \om\ would require knowledge of the host star spectrum, climate and photochemistry modeling, and general atmospheric context. For planets around hotter hosts, another layer of complexity exists, due to the fact that similar \oz\ abundances can occur at very different \om\ values, making it impossible to glean the total \om\ abundance from a measurement of only total \oz\ abundance. However, \oz\ measurements could still provide useful information on the atmosphere and potentially give insight into whether life could exist on the planetary surface. This paper is a continuation of that study, specifically focusing on how \om-\oz\ relationships could vary with different atmospheric compositions.

In this current study, we expand upon the models from \cite{koza22} by varying not only \om\ but also nitrous oxide (\no). Nitrous oxide is particularly interesting in this context, not only because it is biologically produced and considered a promising biosignature (e.g., \citealt{schw18,ange24}), but also because it is the “parent species” of nitrogen oxides (\nox). These nitrogen oxides play a crucial role in the two catalytic cycles that destroy \oz\ as well as the smog mechanism that produces \oz.

Since the rate at which \no\ is converted into \nox\ is dependent on photolysis rates, the degree to which varying \no\ impacts the \om-\oz\ relationship depends on both the overall spectrum of the host star as well as the UV spectral slope. This paper includes all of the models from \cite{koza22}, rerun with different amounts of \no. Section~\ref{sec:chemistry} reviews the relevant chemistry of \oz\ and \no, and Sect.~\ref{sec:methods} describes the atmospheric models, the input stellar spectra, and the radiative transfer model. Section~\ref{sec:results} analyzes how varying \no\ alters atmospheric chemistry, surface UV flux, and simulated planetary emission spectra. Section~\ref{sec:discussions} compares this to similar studies and discusses atmospheric parameters, which affect \no\ abundances in Earth-like planetary atmospheres. A summary of the study and conclusions are available in Sect.~\ref{sec:conclusions}.

\section{Relevant chemistry \label{sec:chemistry}}

\subsection{Ozone formation and destruction \label{sec:oz_formation}}
The majority of \oz\ on Earth is formed via the Chapman mechanism \citep{chap30}, beginning with \om\ photolysis creating atomic O, which then combines with another \om\ molecule with the help of a background molecule, $M$, to carry away the excess energy:
\begin{equation}
    \m{O}_2 + \m{h}\nu \rightarrow \m{O} + \m{O  (}175 < \lambda < 242\ \m{nm}),
    \label{r:PO2_O}
\end{equation}
\vspace{-0.5cm}
\begin{equation}
    \m{O + O}_2 + M \rightarrow \m{O}_3 + M.
    \label{r:O2M}
\end{equation}
\noindent Since Reaction~\ref{r:O2M} is a three-body reaction, it proceeds faster at higher atmospheric densities, with this reaction in particular having a strong temperature dependence, favoring cooling temperatures. While Reaction~\ref{r:PO2_O} creates ground state O atoms (also written as O($^3$P)), the \om\ photolysis initiated by photons with wavelengths less than 175 nm creates the \od\ radical,
\begin{equation}
    \m{O}_2 + \m{h}\nu \rightarrow \m{O }+ \m{O(}^1\m{D)  (}\lambda < 175\ \m{nm}),
    \label{r:PO2_O1D}
\end{equation}
which can then be quenched back to the ground state via collisions with a background molecule,
\begin{equation}
    \m{O(}^1\m{D)} + M \rightarrow \m{O} + M ,
    \label{r:quench}
\end{equation}
or react with other molecules. Similarly \oz\ photolysis creates \om\ and either a ground O atom or an excited \od\ radical depending on the energy level of the photon:
\begin{equation}
    \m{O}_3 + \m{h}\nu \rightarrow \m{O}_2 + \m{O(}^1\m{D)}  (\lambda < \m{310\ nm}),
    \label{r:PO3_O1D}
\end{equation}
\vspace{-0.5cm}
\begin{equation}
    \m{O}_3 + \m{h}\nu \rightarrow \m{O}_2 + \m{O (310} < \lambda < \m{1140\ nm}).
    \label{r:PO3_O}
\end{equation}
After photolysis the resulting O atom often recombines with \om\ via Reaction~\ref{r:O2M}, so the photolysis of \oz\ is not seen as a loss of \oz. Due to the constant cycling between \oz\ and O, it is often useful to keep track of O~+~\oz, termed “odd oxygen," rather than tracking both individually. Odd oxygen can be lost when converted to \om\ molecules, as seen in,
\begin{equation}
    \m{O}_3 + \m{O} \rightarrow 2\m{O}_2.
    \label{r:O3_O}
\end{equation}
The conversion of odd oxygen back into \om\ requires the Chapman mechanism to restart with \om\ photolysis, which is the slowest and limiting reaction of \oz\ formation.
However, Reaction~\ref{r:O3_O} is significantly slower, so O from \oz\ photolysis tends to preferentially combine with \om\ back into \oz\ (Reaction~\ref{r:O2M}). On Earth the majority of \oz\ is created via the Chapman mechanism, with formation rates highest in the stratosphere. This region is high enough in the atmosphere that \om\ is quickly photolyzed, yet still has sufficient atmospheric density for the three-body reaction that creates \oz\ (Reaction~\ref{r:O2M}) to be efficient.

Lower in the atmosphere, primarily in the troposphere, there is another mechanism for \oz\ formation, which is referred to as "smog formation" \citep{haag52}, expressed as,
\begin{equation}
    \m{OH} + \m{CO} \rightarrow \m{H} + \m{CO}_2,
    \label{r:OH_CO}
\end{equation}
\vspace{-0.5cm}
\begin{equation}
    \m{H} + \m{O}_2 + M \rightarrow \m{HO}_2 + M,
    \label{r:H_O2}
\end{equation}
\vspace{-0.5cm}
\begin{equation}
    \m{HO}_2 + \m{NO} \rightarrow \m{OH} + \m{NO}_2,
    \label{r:HO2_NO}
\end{equation}
\vspace{-0.5cm}
\begin{equation}
    \m{NO}_2 + \m{h}\nu \rightarrow \m{NO} + \m{O},
    \label{r:PNO2}
\end{equation}
\vspace{-0.5cm}
\begin{equation}
    \m{O + O}_2 + M \rightarrow \m{O}_3 + M.
    \tag{\ref{r:O2M}}
\end{equation}
\begin{equation*}
\begin{aligned}
\hline
\m{Net:} \hspace{0.5cm}  \m{CO} + 2\m{O}_2 + \m{h}\nu \rightarrow \m{CO}_2 + \m{O}_3
\end{aligned}
\end{equation*}
This chain of reactions requires hydrogen oxides (\hox, \hoo\ + OH + H) and nitrogen oxides (\nox, NO$_3$ + \noo\ + NO) as catalysts, although they are not consumed. On Earth the smog mechanism is slower and produces significantly less \oz\ than the Chapman mechanism, but studies such as \cite{gren13} have demonstrated that planets around cooler spectral hosts may experience much higher efficiency in the smog mechanism. 

While the Chapman and smog mechanisms are the main sources of \oz\ formation, catalytic cycles are the main sources of \oz\ destruction. These are cycles in which there is a net loss of odd oxygen as follows,
\begin{equation*}
\begin{aligned}
\m{X + O}_3 \rightarrow \m{XO + O}_2, \\
\m{XO + O} \rightarrow \m{X + O}_2, \\
\hline
\m{Net:} \hspace{0.5cm}  \m{O}_3 + \m{O} \rightarrow 2\m{O}_2
\end{aligned}
\end{equation*}
with X and XO cycling between each other without loss. On Earth the two most dominant cycles use X = NO and X = OH for the \nox\ and \hox\ catalytic cycles, respectively. The \hox\ catalytic cycle is as follows,
\begin{equation}
\m{OH} + \m{O}_3 \rightarrow \m{HO}_2 + \m{O}_2,
\label{r:HOx_OH}
\end{equation}
\vspace{-0.6cm}
\begin{equation}
\m{HO}_2 + \m{O} \rightarrow \m{OH} + \m{O}_2,
\label{r:HOx_HO2}
\end{equation}
in which odd oxygen, O + \oz, is converted into two \om. In the upper stratosphere where H atoms are more common (often from H$_2$O photolysis) odd oxygen can be converted to \om\ via,
\begin{equation}
\m{H} + \m{O}_2 + M \rightarrow \m{HO}_2 + M,
\label{r:HM}
\end{equation}
\vspace{-0.6cm}
\begin{equation}
\m{HO}_2 + \m{O} \rightarrow \m{OH} + \m{O}_2,
\tag{\ref{r:HOx_HO2}}
\end{equation}
\vspace{-0.6cm}
\begin{equation}
\m{OH} + \m{O} \rightarrow \m{H} + \m{O}_.
\end{equation}
In the lower stratosphere where there is less \om\ photolysis and therefore fewer O atoms, odd oxygen is destroyed via,
\begin{equation}
\m{OH} + \m{O}_3 \rightarrow \m{HO}_2 + \m{O}_2,
\end{equation}
\vspace{-0.6cm}
\begin{equation}
\m{HO}_2 + \m{O}_3 \rightarrow \m{OH} + 2\m{O}_2.
\end{equation}
There are multiple reactions that destroy either OH or \hoo, but they are typically recycled back into another \hox\ species. Photolysis is also not a true sink of \hox, since \hoo\ photolysis creates OH, and OH is too short-lived for significant photolysis. Efficient methods of \hox\ destruction include conversion to H$_2$O,
\begin{equation}
    \m{OH} + \m{HO}_2 \rightarrow \m{H}_2\m{O}+ \m{O}_2,
    \label{r:OH_HO2_H2O}
\end{equation}
or conversion to a stable reservoir species, such as
\begin{equation}
    \m{OH} + \m{OH} + M \rightarrow \m{H}_2\m{O}_2 + M,
    \label{r:OH_OH_H2O2}
\end{equation}
\vspace{-0.3cm}
\begin{equation}
    \m{OH} + \m{NO}_2 + M \rightarrow \m{HNO}_3 + M,
    \label{r:OH_NO2_HNO3}
\end{equation}
\vspace{-0.3cm}
\begin{equation}
    \m{HO}_2 + \m{NO}_2 + M \rightarrow \m{HO}_2\m{NO}_2 + M.
   \label{r:HO2_NO2_HO2NO2}
\end{equation}
The other primary catalytic cycle, the \nox\ catalytic cycle, is discussed in the next subsection, as it is fueled by \no.

\subsection{Nitrous oxide and NO$_x$ catalytic cycles \label{sec:NOx_chem_info}}
Nitrous oxide, considered itself a biosignature, is primarily created by nitrification and denitrification processes in soils and oceans, with very few abiotic sources. On Earth there are additionally many anthropogenic sources of \no, mainly from agricultural processes. The only known major sink of \no\ is photolysis in the stratosphere. \no\ photolysis is one of the sources of  the \od\ radical,
\begin{equation}
    \m{N}_2\m{O} + \m{h}\nu \rightarrow {N}_2 + \m{O(}^1\m{D)},
    \label{r:PN2O}
\end{equation}
which can react with \no\ to create NO,
\begin{equation}
    \m{N}_2\m{O} + \m{O(}^1\m{D)} \rightarrow \m{NO} + \m{NO},
    \label{r:N2O_O1D}
\end{equation}
or N$_2$ and \om,
\begin{equation}
    \m{N}_2\m{O} + \m{O(}^1\m{D)} \rightarrow \m{N}_2 + \m{O}_2.
    \label{r:N2O_O1D_N2_O2}
\end{equation}
While \no\ is the main non-anthropogenic source of \nox, there are additionally minor sources of \nox\ from cosmic rays and lightning (e.g., \citealt{nico75,tuck76,shum03,braam22}), which are not explored in this study. The primary \nox\ catalytic cycle working in the stratosphere follows as,
\begin{equation}
\m{NO} + \m{O}_3 \rightarrow \m{NO}_2 + \m{O}_2,
\label{r:NO_O3}
\end{equation}
\vspace{-0.4cm}
\begin{equation}
\m{NO}_2 + \m{O} \rightarrow \m{NO} + \m{O}_2.
\end{equation}
Nitrate (NO$_3$) is photolyzed at longer wavelengths, allowing NO$_3$ photolysis to occur in the lower stratosphere and further contributing to \oz\ destruction, as follows,
\begin{equation}
\m{NO} + \m{O}_3 \rightarrow \m{NO}_2 + \m{O}_2,
\tag{\ref{r:NO_O3}}
\end{equation}
\vspace{-0.4cm}
\begin{equation}
\m{NO}_2 + \m{O}_3 \rightarrow \m{NO}_3 + \m{O}_2,
\end{equation}
\vspace{-0.4cm}
\begin{equation}
\m{NO}_3 + \m{h}\nu \rightarrow \m{NO} + \m{O}_2.
\end{equation}
\nox\ is lost from the atmosphere through reactions with N atoms, as shown below,
\be
\m{NO} +\m{N} \rightarrow \m{N}_2 + \m{O},
\label{r:NO_N}
\ee
where the N atoms are produced by the photolysis of NO:
\be
\m{NO} +\m{h}\nu \rightarrow \m{N}_2 + \m{O}.
\label{r:PNO}
\ee
Other sinks include conversion to reservoir species, as shown by
\begin{equation}
    \m{NO}_2 + \m{NO}_3 + M \rightarrow \m{N}_2\m{O}_5 + M,
    \label{r:NO2_NO3_N2O5}
\end{equation}
\vspace{-0.3cm}
\begin{equation}
    \m{OH} + \m{NO}_2 + M \rightarrow \m{HNO}_3 + M,
    \tag{\ref{r:OH_NO2_HNO3}}
\end{equation}
\vspace{-0.3cm}
\begin{equation}
    \m{HO}_2 + \m{NO}_2 + M \rightarrow \m{HO}_2\m{NO}_2 + M,
    \tag{\ref{r:HO2_NO2_HO2NO2}}
\end{equation}
in which the reservoir species are significantly less reactive.

\begin{figure}[h!]
    \centering
    \includegraphics[scale=0.32]{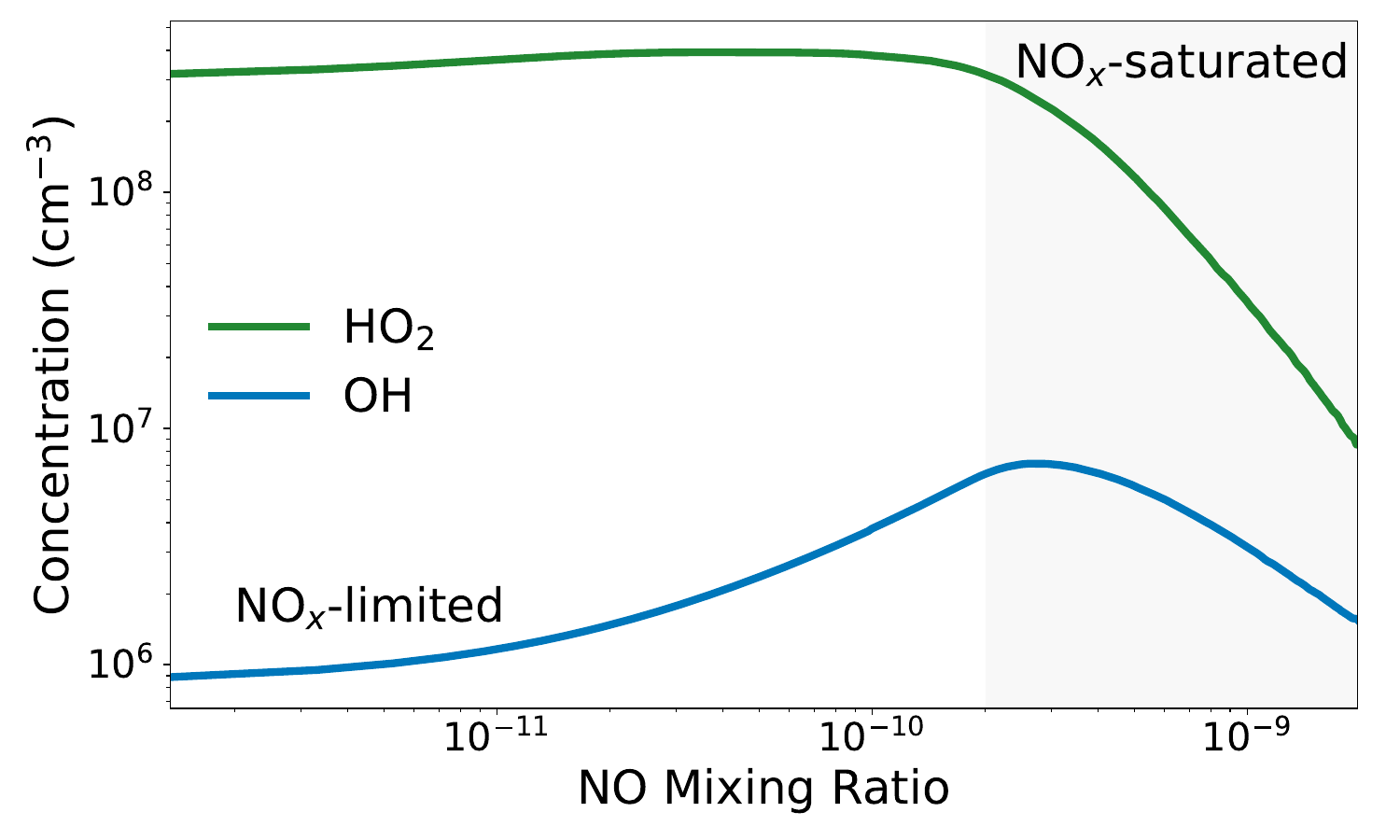}
    \caption{Impact of \hoo\ and OH in the \nox-limited and \nox-saturated regimes on modern Earth, adapted from \cite{loga81}. In the \nox-limited regime (white background), increasing \nox\ allows for a more efficient smog mechanism, with the resulting increase in \oz\ causing a corresponding rise in \hox. In the \nox-saturated regime (gray background), the abundance of \nox\ rises to the point where smog \oz\ production is suppressed as \nox\ depletes \hox\ (a necessary catalyst for the smog mechanism) by locking it up into reservoir species.
    \label{fig:NOx_regimes}}
\end{figure}

\subsection{\nox-limited and \nox-saturated regimes}
Since the smog mechanism uses O atoms created by \noo\ photolysis to create \oz, it might seem that an increase in \nox\ in the lower atmosphere would always lead to a faster smog mechanism.  However, the relationship between \nox\ and the smog mechanism is complicated due to the relationship between \nox\ and \hox. While \nox\ is produced from \no\ in the atmosphere, \hox\ is often indirectly created from \oz\ itself. Hydroxyl (OH) is most commonly created by oxidation of water with the \od\ radical,
\begin{equation}
    \m{H}_2\m{O} + \m{O(}^1\m{D)} \rightarrow \m{OH} + \m{OH},
    \label{r:H2O_O1D}
\end{equation}
with the majority of \od\ in the lower atmosphere created from \oz\ photolysis (Reaction~\ref{r:PO3_O1D}). Therefore, an increase in \oz\ often leads to an increase in \hox. It follows that an increase in \nox\ would allow increased smog production of \oz, with the higher \oz\ concentrations creating more \hox. However, this only holds true in what we call a  “\nox-limited” regime, where increasing \nox\ leads to an increase in \hox. Sufficiently high levels of \nox\ will induce a shift into what we call the  “\nox-saturated” regime, where \nox\ will become efficient at locking up \hox\ into reservoir species such as HNO$_3$ and HO$_2$NO$_2$ (Reactions~\ref{r:OH_NO2_HNO3},\ref{r:HO2_NO2_HO2NO2}). This relationship in the \nox-limited and \nox-saturated regimes as they exist for modern Earth is shown in Fig.~\ref{fig:NOx_regimes}. Since \nox\ and \hox\ are primarily created through reactions of \od\ with \no\ and H$_2$O, respectively (Reactions~\ref{r:N2O_O1D},\ref{r:H2O_O1D}), and \od\ is created via photolysis, the amount of incoming UV from the host star plays a crucial role in determining which \nox\ regime is dominant in the atmosphere.

\begin{figure}[h!]
    \centering
    \includegraphics[scale=0.52]{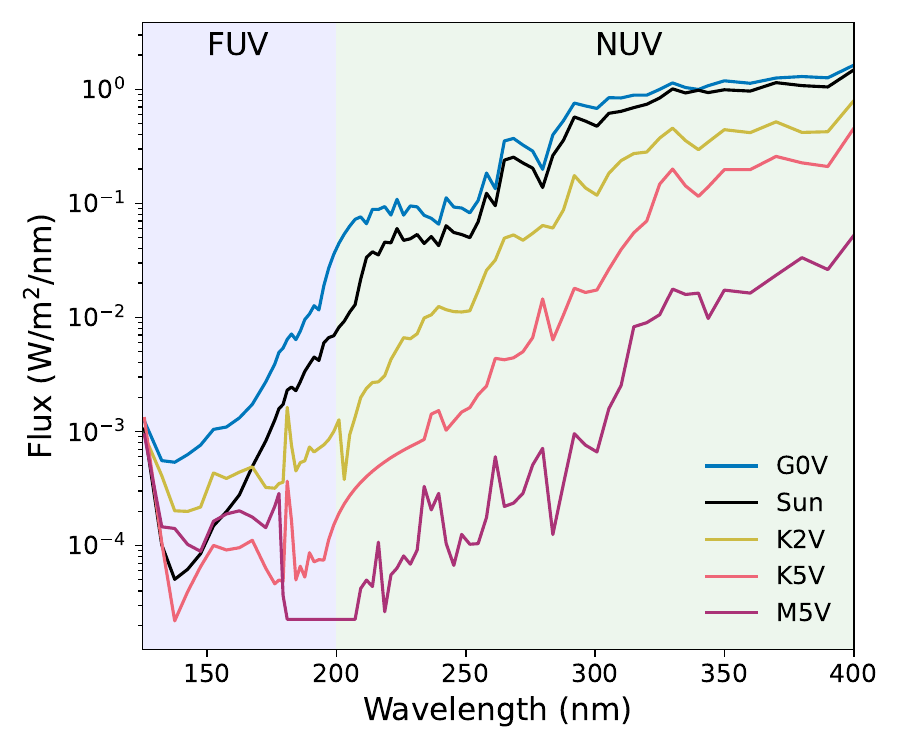}
    \includegraphics[scale=0.52]{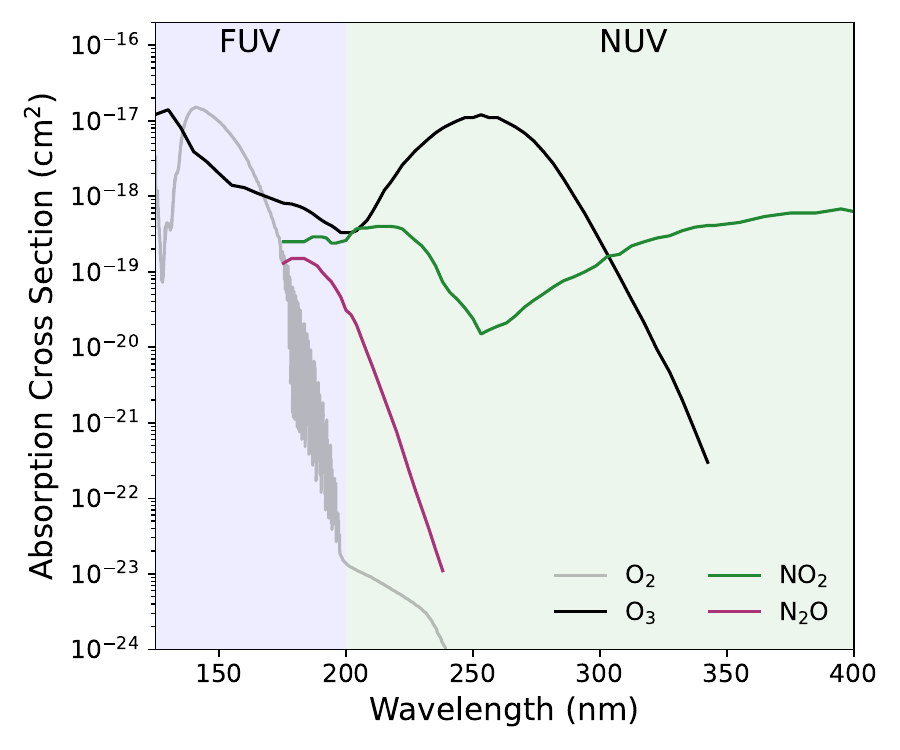}
    \caption{UV stellar spectra of the host stars in this study (top) and corresponding absorption cross sections of relevant gaseous species (bottom). The two plots cover the same wavelength range in order to facilitate comparisons. Cross sections for \noo\ and \no\ are cut off at shorter wavelengths, due to the dominance of absorption from CO$_2$ and other atmospheric species.
    \label{fig:stellar_spectra}}
\end{figure}

\section{Methods \label{sec:methods}}

\subsection{Atmospheric models}
We used \texttt{Atmos}\footnote{https://github.com/VirtualPlanetaryLaboratory/atmos}, a publicly available 1D coupled climate and photochemistry code for atmospheric modeling, following \cite{koza22}. Here, we give a brief summary of the code, with more details available in other papers \citep{arne16,mead18a,koza22}. For inputs \texttt{Atmos} requires a stellar host spectrum (121.6 - 45450 nm), upper and lower boundary conditions for individual gases, initial concentrations of gaseous species, and planetary parameters (radius, gravity, and surface albedo).

We used the modern Earth template for the photochemistry code \citep{kast79,zahn06}, which contains 50 gaseous species and a network of 233 chemical reactions. The atmosphere was calculated up to 100~km and was broken up into 200 plane parallel layers, each solving the flux and continuity equations simultaneously to determine the atmospheric composition. Vertical transport was included for all the species that were not considered “short-lived,” including molecular and eddy diffusion. Radiative transfer calculations used the $\delta$-2-stream method developed in \cite{toon89}, and convergence was reached when the adaptive time step reached the age of the universe ($\sim$10$^{17}$ seconds) within the first 100 steps of the code.

The climate model \citep{kast86,kopp13,arne16} calculates the temperature and pressure profile of the atmosphere based on the incoming stellar radiation and the atmospheric composition. As with the photochemistry code, a $\delta$-2-stream multiple scattering method computes the absorption of stellar flux throughout the atmosphere, and then a correlated-$k$ method computes the absorption of \oz, H$_2$O, CH$_4$, CO$_2$, and C$_2$H$_6$ in each layer for outgoing IR radiation including both single and multiple scattering. The atmosphere was broken up into 100 layers from the surface up until 1~mbar (typically $<$60-70 km), as the code could not be reliably run above these pressures \citep{arne16}. Temperatures and species profiles were held constant above this altitude when transferred to the photochemistry code. Convergence was achieved when the temperature and flux differences out of the top of the atmosphere were deemed sufficiently small ($<$10$^{-5}$) \citep{arne16}.

We ran the climate and photochemistry models, coupled using the “short-stepping” convergence method \citep{teal22,koza22}, iterating back and forth for 30 iterations or until convergence for the two codes was reached. For this study we explored “Earth-like” atmospheres around a variety of host stars, with varying constant mixing ratios of \om\ and \no, building off of \cite{koza22}, who modeled only variations in \om. As in \cite{koza22} we used modern Earth initial conditions for our model atmospheres and varied \om\ from 0.01-150\% PAL. Lower \om\ values were not explored because they are likely unstable \cite{greg21}, and higher values similarly were not explored as they are not compatible with life due to \om\ combustibility \citep{kump08}. All our model planets were run at the Earth-equivalent distance, with a surface pressure of 1~bar, the radius of Earth, and initial gaseous species abundances as listed in the modern Earth \texttt{Atmos} template. This study used all the models from \cite{koza22} with high and low \no\ models using 1000 and 10\% PAL \no, respectively (see Table~\ref{tab:models}). Fixed mixing ratios of \no\ were used (with the modern value of \no=3.0$\times10^{-7}$), to better isolate the effects on \oz. These values were picked in order to begin mapping out the parameter space of atmospheres with different biological fluxes and their impact on \oz\ at different \om\ levels. Since we delved into the \om-\oz\ relationship for changing only \om\ in \cite{koza22}, this paper primarily focuses on how changes in \no\ impact \oz\ levels and the atmosphere generally in comparison to the models with modern levels of \no.

\begin{table}[h!]
\centering
\caption{Model parameters \label{tab:models}}
\begin{tabular}{lcc}
Model name & \no\ MR & \om\ MR \\
\hline
\hline 
\cite{koza22}           & 3.0$\times10^{-7}$    & 2.1$\times10^{-5}$ - 0.315  \\
\hline
Low \no\ (10\% PAL)     & 3.0$\times10^{-8}$ & 2.1$\times10^{-5}$ - 0.315 \\
High \no\ (1000\% PAL) & 3.0$\times10^{-6}$ & 2.1$\times10^{-5}$ - 0.315\\
\hline
\hline
\end{tabular}
\tablefoot{
Abbreviations: MR = mixing ratio; PAL = present atmospheric level.}
\end{table}

\begin{figure*}[h!]
\centering
\includegraphics[scale=0.6]{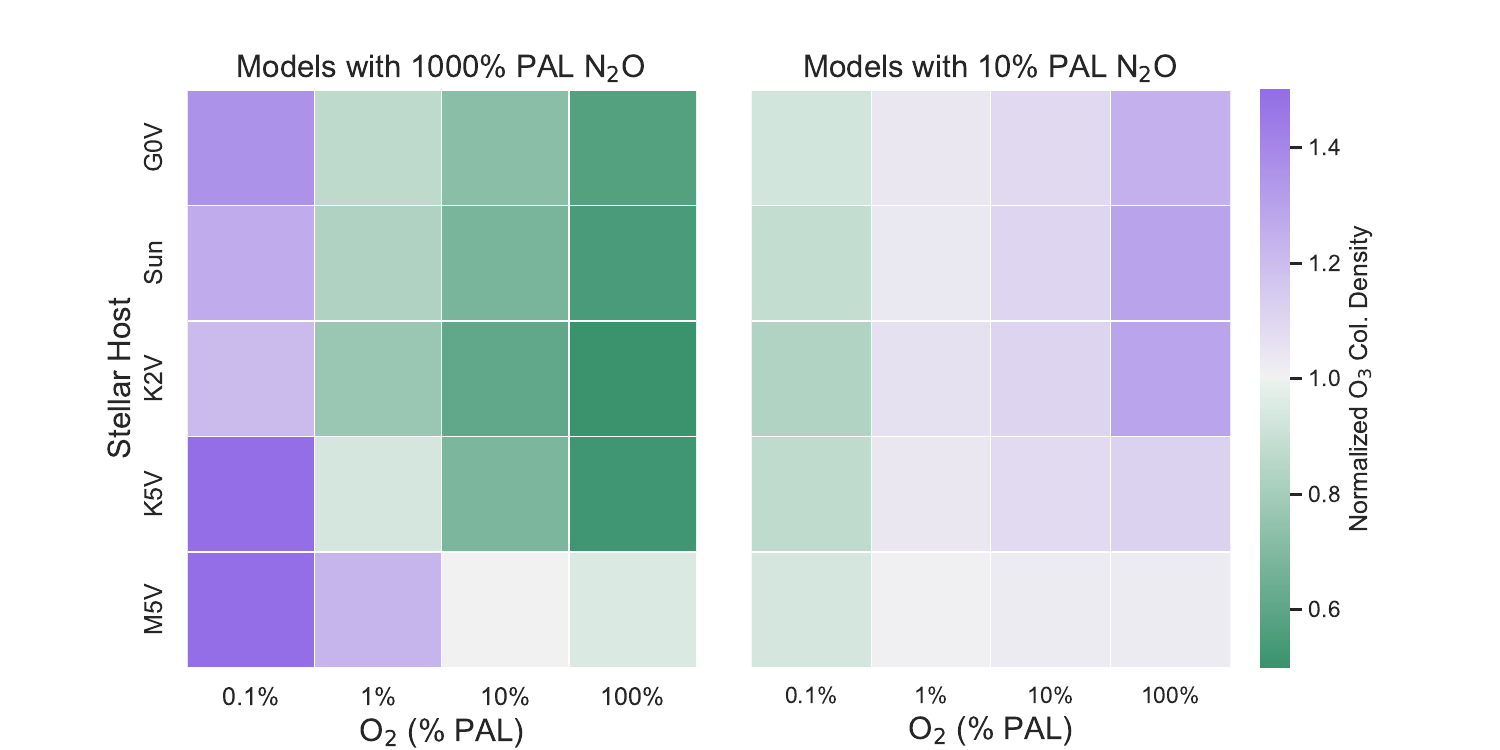}
\caption{Total \oz\ abundances for models with high \no\ (left) and low \no\ (right) normalized to models with modern levels of \no\ from \cite{koza22}. Overall, the high \no\ models impacted the \om-\oz\ relationship more than the low \no\ models, with the results being highly dependent on the stellar host and the amount of \om. High \no\ models with hotter stars experience significant \oz\ depletion due to faster \nox\ catalytic cycles caused by increased \no. However, for the high \no\ models at very low \om\ levels, planets around all the hosts experience an increase in \oz\ due to the higher efficiency of the smog mechanism once the Chapman mechanism is limited by low amounts of \om. The M5V-hosted planet in particular experiences an increase in \oz\ with the high \no\ models begining at 10\% PAL \om\ and lower, due to the increased capabilities of the smog mechanism in this lower UV environment.}
\label{fig:O2O3_relationship_grids}
\end{figure*}

\subsection{Input stellar spectra}
The same \texttt{Atmos} input stellar spectra from \cite{koza22} were used, with hosts ranging from G0V-M5V (for more details, see \citealt{koza22}). The G0V-K5V stellar spectra were originally created in \cite{rugh15} and consist of UV data from \emph{International Ultraviolet Explorer} (IUE) data archives\footnote{http://archive.stsci.edu/iue} combined with ATLAS spectra \citep{kuru79} for the visible and IR wavelength regions. For the M5V star, we used UV data of GJ~876 from the \textit{Measurements of the Ultraviolet Spectral Characteristics of Low-mass Exoplanetary Eystems} (MUSCLES) survey \citep{fran16}. The UV spectra of all host stars are shown in Fig.~\ref{fig:stellar_spectra} along with a comparison to cross sections of gaseous species relevant to \oz\ formation and destruction. The UV spectrum is important for \oz\ formation, particularly the UV spectral slope of the far-UV (FUV; $\lambda <$ 200 nm) and mid- and near-UV (abbreviated NUV, for brevity; 200 nm $< \lambda <$ 400 nm), with later stars tending to have higher FUV/NUV ratios due to activity and high NUV absorption via TiO \citep{harm15}.

\subsection{Post-processing radiative transfer models}
We used the \emph{Planetary Intensity Code for Atmospheric Scattering Observations} (\texttt{PICASO}) to compute planetary emission spectra from our atmospheric models \citep{bata19,bata21}, following \cite{koza22}. This code is publicly available\footnote{https://natashabatalha.github.io/picaso/index.html} with the ability to compute transmission, reflectance, and emission spectra. For emission spectra we input atmospheric profiles of gaseous species and T/P profiles from \texttt{Atmos} and run our models at a phase angle 0$^\circ$ (full phase) for wavelengths of 0.3-14~$\mu$m. We focus in particular on the 9.6~$\mu$m \oz\ feature.

\begin{figure*}[h!]
\centering
\includegraphics[scale=0.5]{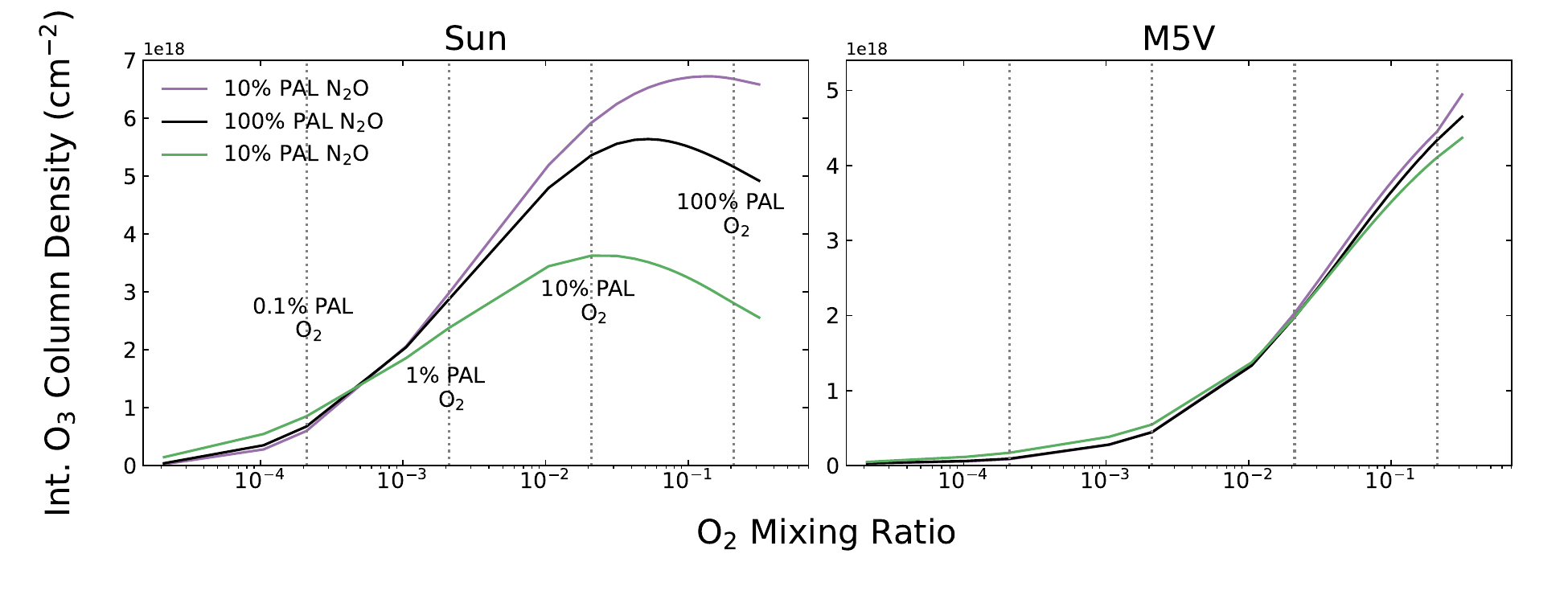}
\caption{\om-\oz\ relationship with high, low, and modern \no\ models for planets around the Sun and M5V hosts. Vertical dashed lines indicate \om\ levels of 100\%, 10\%, 1\%, and 0.1\% PAL. The phenomena causing maximum \oz\ production for the planet around the Sun to occur at \om\ values less than the maximum \om\ value modeled are explained in depth in \cite{koza22}. In general, the M5V-hosted planet experiences significantly less variation in \oz\ for different \no\ abundances than the Sun-hosted planet because the low UV flux of the M5V host is not as efficient at converting \no\ into \nox\ as other hosts with higher UV fluxes. \om-\oz\ relationships for planets around all hosts and comparisons to \cite{koza22} are in Figure~\ref{fig:O2O3_relationship}.}
\label{fig:Sun_M5V_O2O3}
\end{figure*}

\section{Results \label{sec:results}}
Here, we explore the results of our planetary atmospheric models for high and low \no\ and compare them to the \cite{koza22} atmospheric models with modern levels of \no. The impact of varying \no\ levels on the \om-\oz\ relationship is shown to be nonlinear in all cases, with a strong dependency on both the stellar host and the amount of atmospheric \om. In Sect.~\ref{sec:N2O_chem} we analyze the atmospheric chemistry of the varying \no\ models, the variations in UV to the ground in Sect.~\ref{sec:UV_to_ground}, and the resulting planetary emission spectra in Sect.~\ref{sec:emission_spectra}. Additional supporting figures and tables are available in Appendix~\ref{append:A}.

\subsection{Atmospheric chemistry \label{sec:N2O_chem}}

\subsubsection{\emph{Atmospheric chemistry}: Overview}
Changing \no\ is shown to have varying effects on the \om-\oz\ relationship depending on the \om\ level and stellar host, with stronger effects seen with high \no\ models when compared to the low \no\ models. Results for \oz\ abundances normalized to modern levels of \no\ with the high and low \no\ models at 100\%, 10\%, 1\%, and 0.1\% PAL \om\ for all hosts are in Fig.~\ref{fig:O2O3_relationship_grids}, with absolute \oz\ values for all \om\ levels modeled for the Sun- and M5V-hosted planets in Fig.~\ref{fig:Sun_M5V_O2O3}. Absolute \oz\ results for planets around all hosts at all modeled \no\ and \om\ levels with a comparison to results from \cite{koza22} are located in the Appendix (Fig.~\ref{fig:O2O3_relationship}). For planets around all hosts at \om\ levels near 100\% PAL there is a decrease in \oz\ for high \no, and an increase in \oz\ for low \no. Planets hosted by all stars except the coolest one (M5V) experience a large amount of \oz\ depletion at \om\ levels similar to modern Earth for the high \no\ models, with K2V having the overall largest depletion at 100\% PAL, only retaining 47\% of its \oz\ when compared to models with modern levels of \no. In contrast for the corresponding model for the M5V-hosted planet 95\% of the original \oz\ remains.

For the low \no\ models at 100\% PAL \om\ the planets around both the Sun and the K2V host are most impacted, with \oz\ abundances of 129\% of the \oz\ they had with modern \no\ levels.  Again, the planet around the M5V host is least affected, with 103\% of the \oz\ compared to results from modern \no\ levels. At low \om\ levels for the high \no\ models planets around all hosts experience an increase in \oz, with this effect being most significant for the planet around the M5V host. The main factors at work determining the impact of \no\ on \oz\ abundance are:
\begin{itemize}
    \item the balance between the host star's ability to convert \no\ into \nox\ and to destroy \no\ via photolysis
    \item whether the amount of \nox\ reaches the threshold to enter the \nox-saturated regime, which inhibits the smog mechanism
\end{itemize}
Both of these concepts are discussed at length in the following subsections.

\subsubsection{\emph{Atmospheric chemistry}: Efficiency of conversion of \no\ into \nox\ and \no\ photolysis}
The degree to which varying \no\ impacts \oz\ abundance is largely dependent on the host stars' ability to convert \no\ into \nox, as well as the rate at which \no\ is destroyed via photolysis. Although  \no\ levels are the same for planets around all host stars, the rate at which the incoming host star flux converts \no\ to \nox\ varies based on the UV spectrum of the host. Conversion of \no\ into \nox\ requires an \od\ radical (Reaction \ref{r:N2O_O1D}), which is created via,
\begin{equation}
\m{O}_2 + \m{h}\nu \rightarrow \m{O }+ \m{O(}^1\m{D)  (}\lambda < 175\ \m{nm}),
\tag{\ref{r:PO2_O1D}}
\end{equation}
\vspace{-0.6cm}
\begin{equation}
\m{O}_3 + \m{h}\nu \rightarrow \m{O}_2 + \m{O(}^1\m{D)  (}\lambda < 310\ \m{nm}),
\tag{\ref{r:PO3_O1D}}
\end{equation}
\vspace{-0.6cm}
\be
\m{N}_2\m{O} + \m{h}\nu \rightarrow \m{N}_2 + \m{O(}^1\m{D) (}\lambda < 200 \m{ nm)},
\label{r:PN2O_O1D}
\ee
\vspace{-0.6cm}
\be
\m{CO}_2 + \m{h}\nu \rightarrow \m{CO} + \m{O(}^1\m{D) (}\lambda < 167 \m{ nm)}.
\label{r:PCO2_O1D}
\ee
As all of these reactions require short-wavelength UV photons, naturally planets with hotter hosts and higher UV fluxes are more efficient at creating \od, and therefore at converting \no\ into \nox. However, the high UV fluxes of these hosts also destroy significant amounts of \no\ via photolysis -- the main sink of \no. This depletion of \no\ via photolysis becomes even more significant for lower \om\ levels, as photolysis in general can occur deeper in the atmosphere when there is less UV shielding from \om\ and \oz. This effect is discussed at length in \cite{koza22}. The end result is that although the G0V and Sun hosts are most efficient at converting \no\ into \nox, the high levels of \no\ destruction from photolysis mean that \nox\ production is hindered, especially at low \om\ levels, due to the loss of the source \no. The host star displaying the largest increase in \nox\ with the high \no\ models is the K2V host, as it exists in a “sweet spot” where the UV flux is capable of creating enough \od\ to fuel conversion of \no\ into \nox, but not enough UV for significant \no\ depletion to hinder \nox\ production. Planets around the M5V host show the least variation in \nox\ production when varying \no, due to low incoming UV flux.

When varying \no\ the amount of \nox\ created is the main driver of stratospheric \oz\ destruction via the \nox\ catalytic cycle, resulting in overall depletion of \oz\ for high \no\ models and increase in \oz\ for low \no\ models in the majority of cases. However, for our lowest \om\ levels ($\sim$0.1\% PAL) and the majority of M5V-hosted models, the reverse is shown. This is due to the smog mechanism of \oz\ production.

\subsubsection{\emph{Atmospheric chemistry}: Smog mechanism efficiency and \nox-limited and -saturated regimes}
When varying the abundance of \no\ -- and therefore \nox\ -- the smog mechanism of \oz\ production becomes particularly relevant. Described in Sect.~\ref{sec:oz_formation}, the smog mechanism uses \nox\ and \hox\ as catalysts to produce \oz\ in the lower atmosphere. This mechanism sources O atoms for \oz\ formation from \noo\ photolysis, rather than \om\ photolysis as with the Chapman mechanism. The ability for \noo\ to be photolyzed by photons spanning the NUV and into the visible spectrum is in stark contrast to 242~nm required for \om\ photolysis (see Fig.\ \ref{fig:stellar_spectra}). Compared to the Chapman mechanism the smog mechanism can take place much deeper in the atmosphere and without the strong dependence on the host stars' NUV flux. Although smog \oz\ production is increased for all high \no\ cases, for the G0V-K5V hosted planets with \om\ levels of $\sim$1\% PAL and higher the destruction of \oz\ via the \nox\ catalytic cycle outweighs the extra smog-produced \oz\ in the troposphere. However, at 0.1\% PAL \om\ all cases with high \no\ models show an increase in overall \oz\ abundance when compared to models with modern levels of \no. This is because for low \om\ levels the Chapman mechanism becomes severely limited by the amount of \om, while the smog mechanism is less impacted as it relies on \noo\ photolysis instead. However, increased \oz\ smog production resulting in high \oz\ abundances is evident for the M5V-hosted planet at much higher \om\ levels than other hosts, starting at 10\% PAL \om.

\begin{figure}
\centering
\includegraphics[scale=0.35]{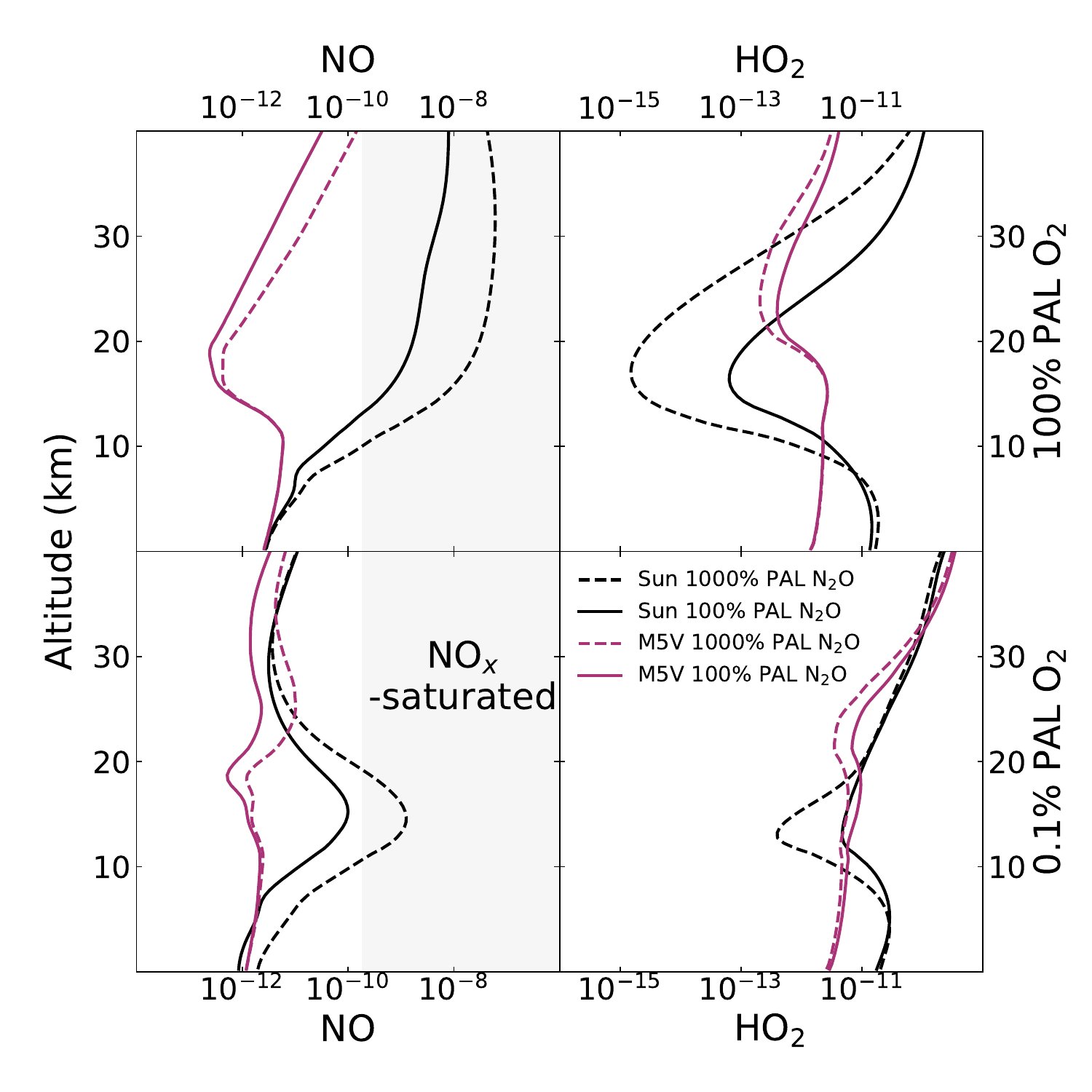}
\caption{NO and \hoo\ profiles at 100 and 0.1\% PAL \om\ for planets around the Sun and M5V hosts. Plots containing NO profiles indicate the \nox-saturated regime by a gray background, while the \nox-limited regime has a white background. When NO profiles enter the \nox-saturated regime, a corresponding depletion of \hoo\ appears in the same part of the atmosphere. The Sun-hosted planet often enters the \nox-saturated regime, due to the efficient conversion of \no\ into \nox. By contrast, the M5V-hosted planet is less efficient at creating \nox\ in its low UV environment and remains in the \nox-limited regime.
\label{fig:NO_HO2}}
\end{figure}

There are two main reasons why the M5V-hosted planets experience a much stronger increase in smog mechanism \oz\ compared to other hosts:
\begin{itemize}
    \item the smog mechanism is much more accessible than the Chapman mechanism with lower incoming UV, as it is easier for the flux of the M5V host to photolyze \noo\ rather than \om\ 
    \item only planets hosted by the M5V star never enter the \nox-saturated regime, meaning that the smog mechanism is not suppressed 
\end{itemize}
This concept of \nox\ regimes (discussed in Sect.~\ref{sec:NOx_chem_info}  and demonstrated in Fig.~\ref{fig:NOx_regimes}) is well-illustrated in Fig.~\ref{fig:NO_HO2}, in which NO and \hoo\ mixing ratios for the Sun- and M5V-hosted planets are compared. Significant depletion of \hoo\ is observed in parts of the atmosphere existing in the \nox-saturated regime for the planet around the Sun. In the \nox-limited regime, increasing \nox\ allows the smog mechanism to create more \oz, and \oz\ in turn creates \od\ radicals that create \hox. However, when \nox\ levels are high enough to be in the \nox-saturated regime \nox\ is efficient at removing \hox\ by locking it up in reservoir species (such as HNO$_3$ and HO$_2$NO$_2$), so an increase in \nox\ leads to a decrease in \hox. As hotter stars are more efficient at converting \no\ into \nox\ -- with the highest efficiency at high \om\ -- NO levels become high enough to enter the \nox-saturated regime and significantly reduce the effectiveness of the smog mechanism. For G0V-K2V hosted-planets with 100\% PAL \om\ all levels of \no\ explored in this study sustain high enough NO mixing ratios in parts of the lower atmosphere to be in the \nox-saturated regime. At modern and high \no\ levels parts of the atmosphere also enter the \nox-saturated regime for the G0V-, Sun-, and K2V- hosted planets at 10\% PAL \om, as well the G0V and Sun cases slightly at 1\% PAL \om. On the other hand, the low UV flux of the M5V host star struggles to convert \no\ into \nox, consistently keeping it in the \nox-limited regime and allowing for a boost in smog \oz\ production whenever \no\ is increased.

\begin{figure*}
\centering
\includegraphics[scale=0.43]{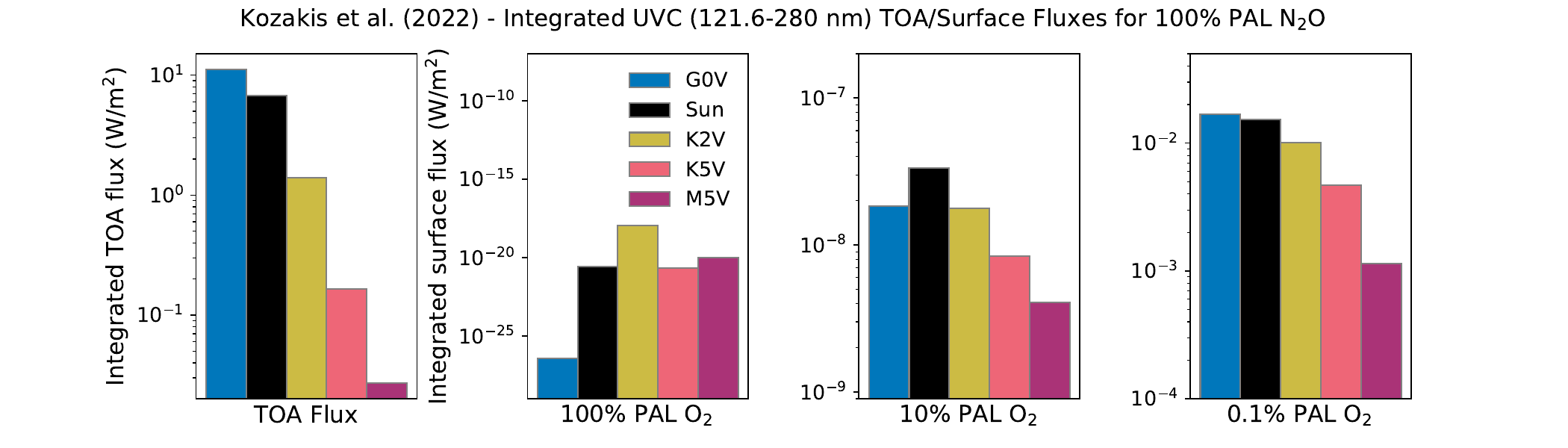}\\
\includegraphics[scale=0.43]{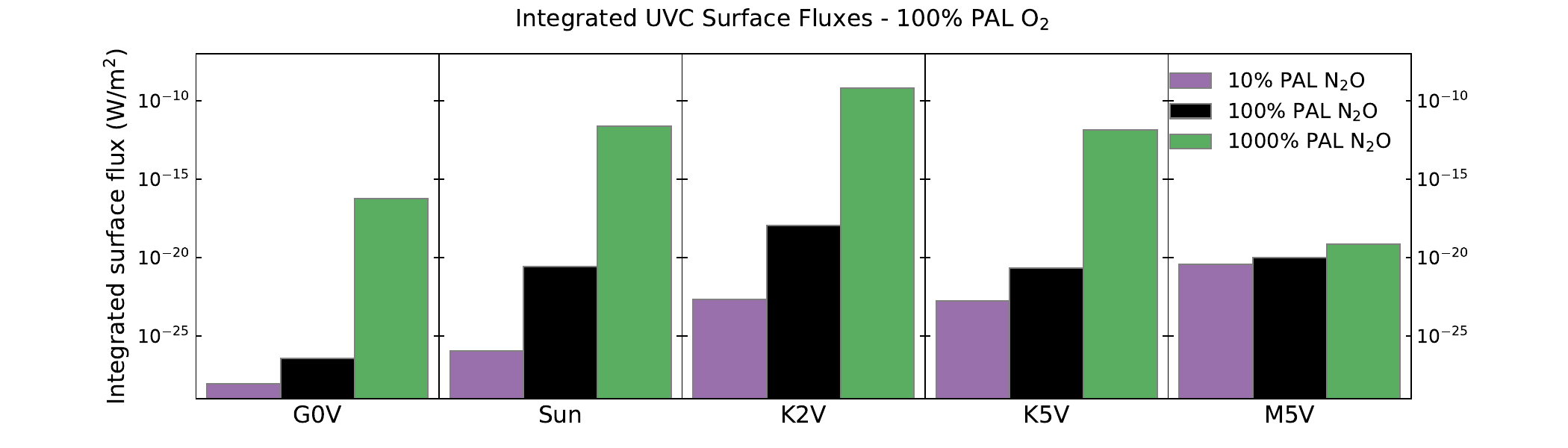}\\
\includegraphics[scale=0.43]{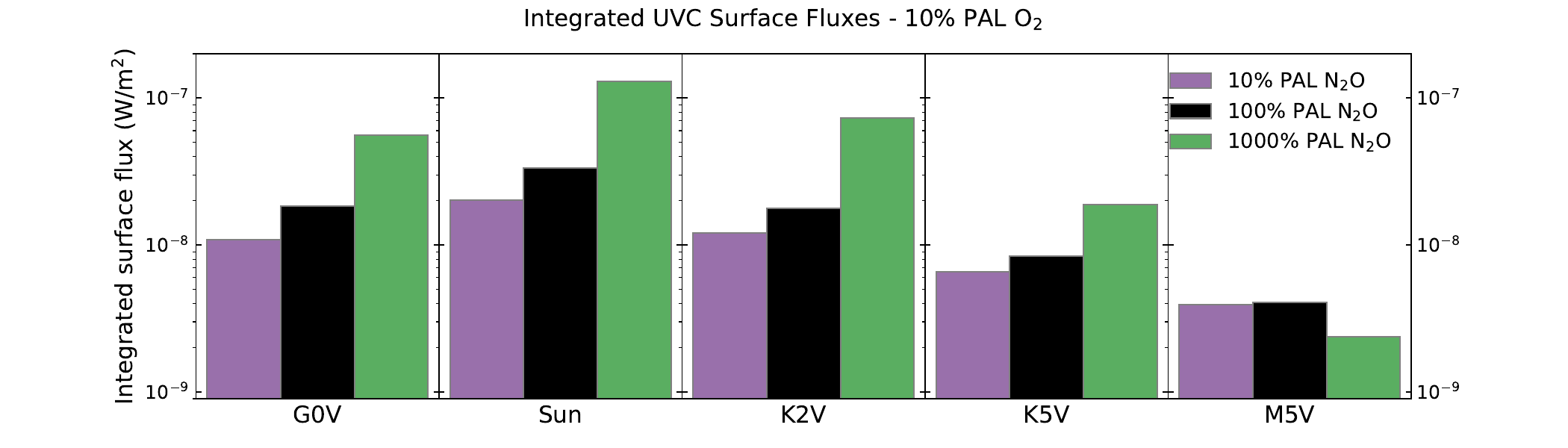}\\
\includegraphics[scale=0.43]{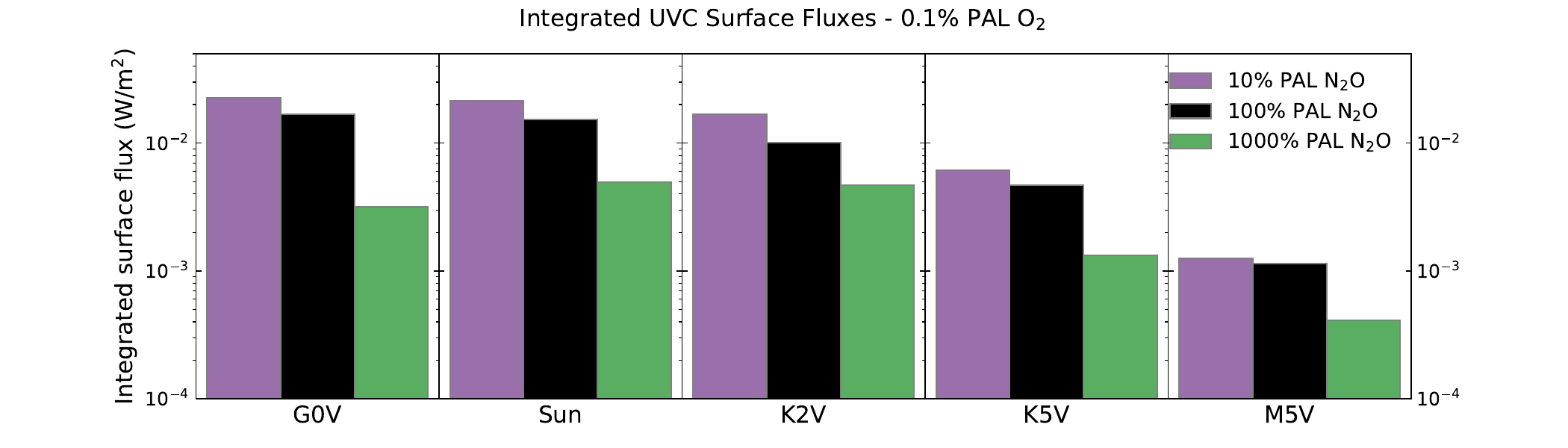}\\
\caption{UVC surface fluxes for planets around all hosts from \cite{koza22} with modern \no\ (top) and with high and low \no\ abundances from this study with \om\ levels of 100\% (second row), 10\%, (third row), and 0.1\% (bottom) PAL. The first panel in the top row indicates the top-of-the-atmosphere (TOA) flux for planets around all host stars, with the rest of the top row using the same y-axis limits for the corresponding \om\ values used in the bottom three rows to facilitate comparisons. Results for 1\% PAL \om\ are extremely similar to those for 10\% PAL \om\ and are therefore not included. The most significant changes in UVC surface flux occur in models with higher \om\ because the conversion from \no\ to \nox\ is more efficient in higher \om\ environments. Larger amounts of \nox\ have a greater impact on \oz\ either by depleting it with \nox\ catalytic cycles or producing it with the smog mechanism. While the results at 10\% and 100\% PAL \om\ tend to show increased \oz\ shielding for low \no\ models, this trend reverses at 0.1\% PAL \om, where the low \no\ models consistently receive more surface UVC. This is because at low \om\ levels the effects of the smog mechanism become clearer as the Chapman mechanism is limited by the lack of \om.}
\label{fig:UVC_barchart}
\end{figure*}

\subsection{UV to ground \label{sec:UV_to_ground}}
This section explores how the changes in the \om-\oz\ relationship translate to changes in the surface UV environment when varying \no. On modern Earth both \om\ and \oz\ are very important for UV shielding, which is important for surface life to flourish. Although \om\ is not as efficient at shielding UV as \oz\ (see a comparison of absorption cross sections in Fig.~\ref{fig:stellar_spectra}), it is significantly more abundant than \oz, and thus being a large constituent of our atmosphere it provides significant shielding. We discuss surface UV environments using three biological regimes of UV: UVA (315-400~nm) is least damaging and may help power complex processes necessary for life; UVB (280-315~nm), which is more dangerous and has been linked to tanning of skin as well as skin cancer; and UVC (121.6-280~nm), which is dangerous for biological organisms and can break apart DNA. On modern Earth UVA is not largely shielded by \oz\ or \om, UVB is partially shielded by \oz, and fortunately UVC is almost entirely shielded by \om\ and \oz, protecting surface life. UVB and particularly UVC have a nonlinear relationship with the amount of \om/\oz, and are very sensitive to changes in \oz.  See \cite{koza22} for an in-depth description of the impacts of UV on the ground while varying only \om. Comparisons of the top-of-the-atmosphere (TOA) and integrated UVC surface flux (the most variable results) are shown in Fig.~\ref{fig:UVC_barchart}, with an additional table in the appendix (Table~\ref{tab:UV_all}) displaying absolute and normalized UVB and UVC surface fluxes. UVA results were unaffected when changing \no, with all models allowing $\sim$80\% of incoming UVA to reach the surface, as in \cite{koza22}. This is unsurprising as \oz\ does not provide shielding at these wavelengths, with other species causing minimal absorption.

UVB surface flux displays more variation when changing levels of \no, but always within an order of magnitude. Higher \om\ levels allow larger variations in UVB surface fluxes, as there are typically higher \oz\ levels, resulting in greater absolute changes in \oz\ abundance and UV shielding ability. The high \no\ models showed the largest change in surface UVB flux at high \om\ levels as increased efficiency of the \nox\ catalytic cycle caused significant changes in \oz\ for planets around all hosts except for the M5V. The most variation was at 100\% PAL \om\ for the G0V-hosted planet with 1.8 times as much UVB flux reaching the surface for the high \no\ models, and 0.72 times as much flux reaching the surface for low \no\ models. The only planet not experiencing large changes in UVB surface flux at these \om\ levels for varying \no\ cases was the one hosted by the M5V star, as the low UV flux did not allow efficient conversion of \no\ into \nox, and  did not impact \oz\ abundance as much as other hosts.

The most significant changes in the UV surface environment were for the UVC surface flux as it covers the wavelength range in which \oz\ shielding is most effective, causing \oz\ changes from varying \no\ to strongly impact UVC surface flux. Similarly to UVB surface flux, for all but the planet hosted by the M5V star, the largest changes in UVC surface flux were caused by the high \no\ cases at higher \om\ levels. This was due once again to the significantly increased efficiency of the \nox\ catalytic cycle with increased \nox\ sourced from \no. The largest increase was observed for the G0V-hosted planet with the high \no\ model at 100\% PAL \om, which experienced an astounding 15 billion-fold increase in UVC surface flux. However, even with this extreme increase in surface UVC flux this case still experiences less surface UVC flux than planets around all other hosts except the M5V for the high \no\ models at this \om\ level. This is because initially the G0V-hosted planet had the strongest UVC shielding at 100\% PAL \om\ due to having the highest \oz\ abundance. The ability of the high \no\ models to deplete \oz\ with sufficient \om\ is still seen at the 10\% PAL \om\ level where there were still significant increases in surface UVC flux for planets around all hosts except for the M5V, which had less surface UVC due to increased amounts of smog produced \oz\ around hosts with lower UV. For our lowest \om\ levels planets around all hosts start to show decreased levels of UVC surface flux for the high \no\ models compared to modern levels of \no\ due to the extra UV shielding from the \oz\ produced by the smog mechanism. 

The largest reduction of UVC surface flux from the low \no\ models was observed for the Sun-hosted planet, with a factor of 4.1$\times10^{-6}$ times the original UVC surface flux with modern \no\ at 100\% PAL \om, due to a less productive \nox\ catalytic cycle. From the high \no\ models the largest UVC surface flux reduction was observed for the planet around the G0V host at 0.1\% PAL \om, with only a factor of 0.19 UVC flux reaching the surface compared to modern \no\ levels due to extra \oz\ from smog production. Overall the \no\ models with the least UVC surface flux variation were those around the M5V host, due largely to the fact that these cases had the least amount of total \oz, so the absolute amount of surface flux does not have much variation.

\begin{figure*}
\centering
\includegraphics[scale=0.5]{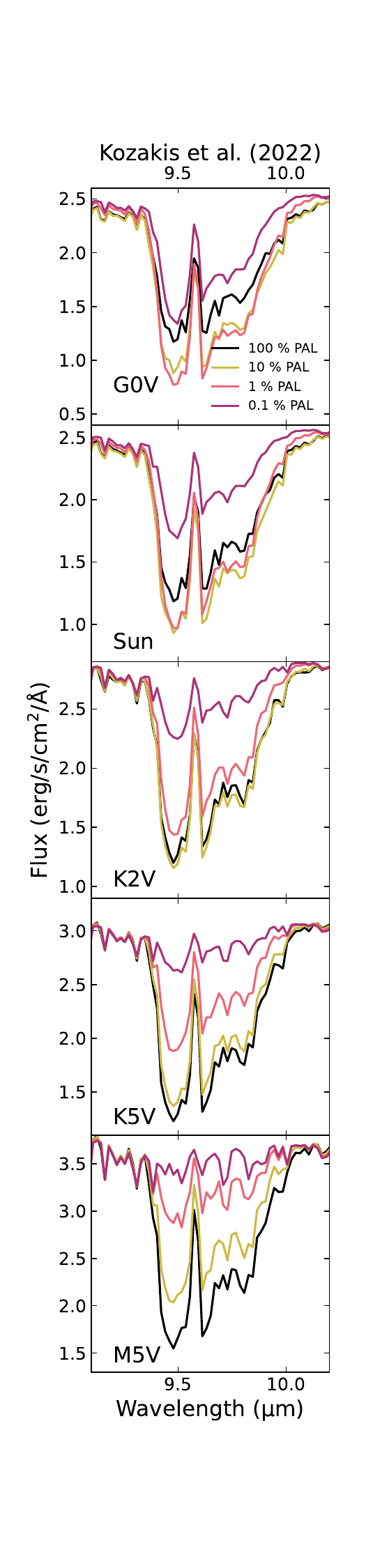}
\hspace{-1.2cm}
\includegraphics[scale=0.5]{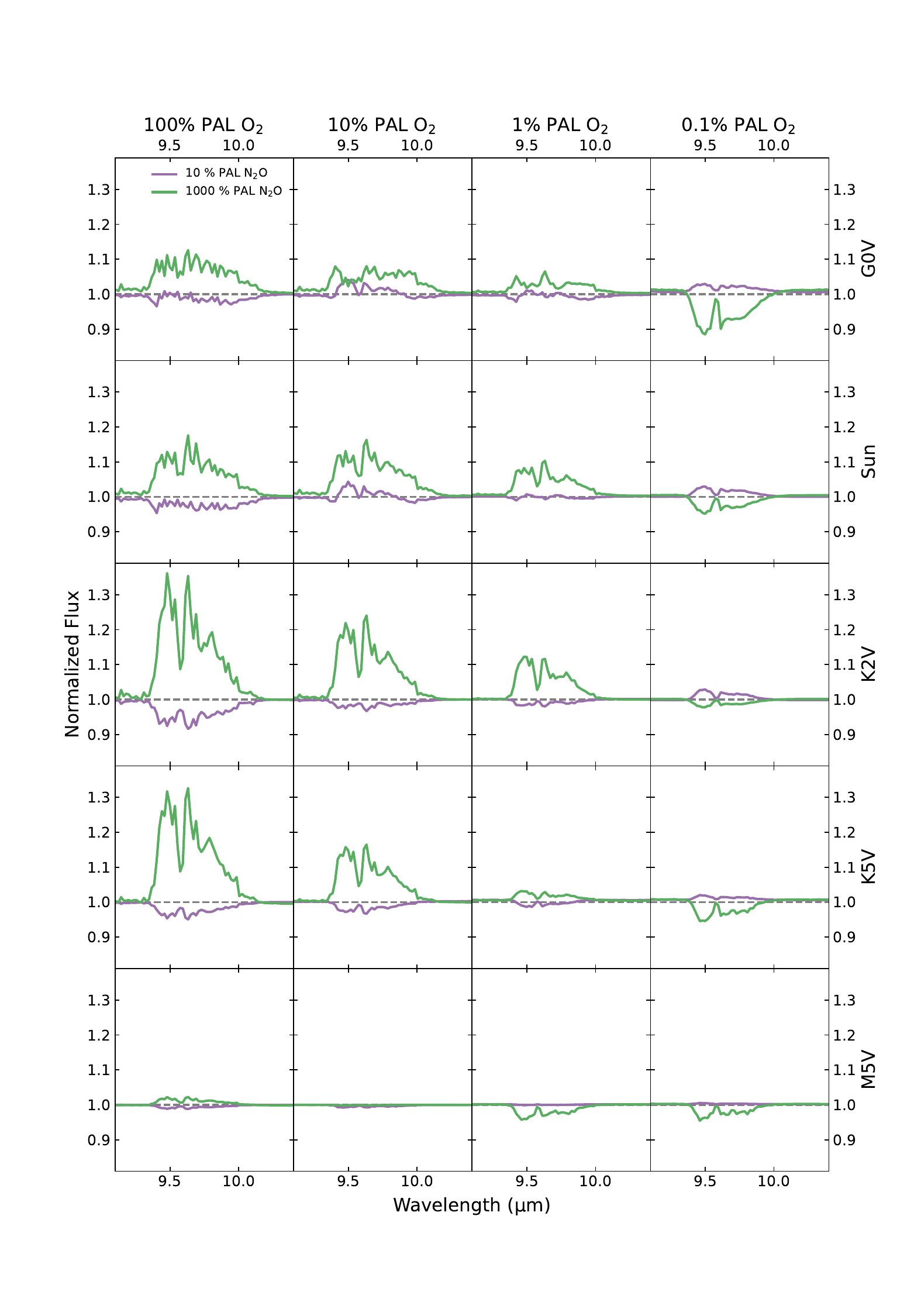}
\caption{9.6 $\mu$m \oz\ feature for models from \cite{koza22} (left) and normalized models from varying \no\ (right) from simulated planetary emission spectra. Changes in the \oz\ feature for this study depended primarily on changes in \oz\ abundance -- rather than on changes in the atmospheric temperature profile, as in \cite{koza22} -- 
since varying \no\ often did not significantly impact the temperature profiles. As the K5V-hosted planet experienced the greatest change in \oz\ abundance with the high \no\ models, it follows that its \oz\ emission spectral features were the most impacted. The M5V-hosted planet experienced the smallest change in \oz\ while varying \no, which is reflected by the small variations in its \oz\ feature.
}
\label{fig:emission}
\end{figure*}

\subsection{Planetary emission spectra \label{sec:emission_spectra}}
In this section we examine how the effects of \no\ on the \om-\oz\ relationship would impact potential future observations. We focus in particular on the 9.6~$\mu$m \oz\ feature to see how it would change for different \no\ abundances. Already in \cite{koza22} we found that just varying levels of \om\ results in counterintuitive changes in the depth of the \oz\ spectral feature. This is due to the fact that spectral feature depth for emission spectra is dependent on not only the abundance of \oz, but also the temperature difference between the absorbing and emitting layers of the atmosphere. This causes a nonlinear relationship between \oz\ abundance and feature depth as \oz\ has a significant impact on stratospheric heating. Ozone absorption features for planets with significant stratospheric heating will be shallower due to a decreased temperature difference between the stratosphere and surface when compared to a planet with less \oz\ and less stratospheric heating. This effect impacts planets around hotter stars more than those around cooler stars, as \oz\ heats the atmosphere via absorption of NUV photons and cooler stars (especially M dwarfs) have less NUV flux due lower temperatures and TiO absorption. Since the atmospheres of planets around cooler stars tend to be more isothermal, spectral feature depth has a more linear relationship between \oz\ abundance and \oz\ feature depth, as seen in the left-hand side of Fig.~\ref{fig:emission}, which displays emission spectra from \cite{koza22}. For a more in-depth look at how this effect impacts \oz\ spectral features with varying \om\ levels please refer to \cite{koza22}.

Figure~\ref{fig:emission} shows 9.6~$\mu$m \oz\ features with varying \no\ normalized to features with modern levels of \no\ in order to understand how observations of \oz\ could be impacted by varying \no\ abundances. When comparing model atmospheres with varying \no\ to those with modern levels of \no, the largest changes in temperature profiles (and, therefore, the spectral features) were due to changing amounts of stratospheric heating from \oz, which often was not a large change. As a result, changes in the \oz\ feature depth were due primarily to variations in \oz\ abundance, rather than the atmospheric temperature profile as in \cite{koza22}. Overall the most significant changes in spectral feature strength were caused by the high \no\ models at higher \om\ levels, which is unsurprising as these are the cases with the largest depletion in \oz\ from the enhanced \nox\ catalytic cycle. The planet around the K2V host at 100\% PAL \om\ with the high \no\ model has the spectral feature that changes the most compared to modern levels of \no, with a shallower feature caused by \oz\ depletion. At 0.1\% PAL \om\ for planets around all hosts there is a deeper \oz\ feature due to the extra \oz\ produced by the smog mechanism at low levels of \om. There is less variation in spectral feature strength for the low \no\ models as they had a much weaker effect on \oz.

\section{Discussion \label{sec:discussions}}

\subsection{Comparisons to other studies}
Here we discuss studies examining how changes in \no\ would impact a planet, especially with different \om\ and \oz\ levels. For a full review of studies exploring the relationship between \om\ and \oz, see \cite{koza22}. While no other study varied \no\ specifically to look at changes in the \om-\oz\ relationship, there exist studies similar enough that it is useful to compare trends.

\cite{gren06} explores the possibility that on early Earth during periods with low \om, \oz\ produced by the smog mechanism using \noo\ photolysis instead of \om\ photolysis could have provided UV shielding for surface life. Motivated by the fact that \hox\ levels and possibly \no\ levels were higher during the Proterozoic period (2.4-0.54 Gyr ago), they varied \ch, \om, \nox, H$_2$, and CO abundances to study the impact on smog \oz\ formation using a photochemistry box model. Although they explore a different parameter space than in this study, similar trends appear, and they also see the effects of the atmosphere in a \nox-saturated regime, which suppresses \oz\ formation, similarly to what we find at high \no\ in our model atmospheres around the hotter stars.

There are also several studies in a similar vein focusing on planets in M dwarf systems, especially since the low incoming UV from such hosts could lead to a buildup of \no\ in their atmospheres \citep{segu03,segu05}. \cite{raue11} and \cite{gren13} -- both part of the same paper series -- discuss the increased smog production efficiency of planets around M dwarfs, particularly late-type M dwarfs. This idea is explored in depth in \cite{gren13}, where they use the Pathway Analysis Program \citep{lehm04} to compare the efficiencies of the Chapman and smog mechanisms of \oz\ production. Although they use an earlier version of \texttt{Atmos} for their atmospheric modeling, they utilize a more complex chemical network to study smog formation, including variations of the smog mechanism that are not included in our chemical network that involve more complex methyl-containing molecules (e.g., CH$_3$O$_2$). However, the “classical” smog mechanism that we use in this study is shown to be the most common type of smog production. The trends reported in \cite{gren13} agree with those discussed in this study, especially with our results for increased smog production of \oz\ around our coolest host, the M5V. However, \cite{gren13} do not vary \om\ or \no\ abundances. Another study, \cite{gren14}, varies \no\ and \ch\ biological surface fluxes, along with incoming UV fluxes to explore the effect of a planet orbiting an M7V host star. Although they use an M7V host star and explore different \no\ abundances than in this paper (\no\ at a factor of 1000 lower and zero \no\ flux), similar trends are observed -- particularly decreased smog production for decreased \no, which we see especially in our models around cooler host stars.

\cite{schw22} explores what a plausible range of \no\ surface fluxes could be for terrestrial planets using both a biogeochemical model and a photochemistry model (from \texttt{Atmos}). Using the biogeochemical model cGENIE they predict possible \no\ surface fluxes based off denitrification process at \om\ abundances from 1-100\% PAL with planet hosts ranging from F2V to M8V. By modeling total ocean denitrification they determine maximum \no\ atmospheric mixing ratios for different \om\ abundances. Their results show that our high \no\ models (using an \no\ mixing ratio of 3 ppm) are safely within the range of possible \no\ values for all spectral host types. Although \cite{schw22} also explores atmospheres with different amounts of \no\ and \om\ as in our study, the focus is not on \oz\ so it is difficult to make direct comparisons. \cite{schw22} briefly show that increasing \no\ results in increased destruction of stratospheric \oz, agreeing with our general results, although they do not discuss if increased smog mechanism formation occurs at large \no\ values. Overall, when comparing our current study to other similar works there appears to be no real inconsistencies, although the studied parameter spaces are varied enough that it is not possible to make direct comparisons.

\subsection{Plausible \no\ mixing ratios in Earth-like atmospheres}
In this study we used fixed mixing ratio profiles for \no\ to facilitate comparisons of how those levels of \no\ were expected to impact the \om-\oz\ relationship across different host stars. However, it is important to recognize that the relationship between surface flux and resulting atmospheric mixing ratios is not linear, and is highly influenced by the host star as well as atmospheric content. For \no\ in particular the \om\ level has significant impact on the resulting mixing ratio, meaning that there would be a range of \no\ fluxes that would be necessary to reproduce the mixing ratios used in this study. As stated in Sect.~\ref{sec:methods}, the motivation behind our chosen \no\ abundances is to begin filling in the parameter space of how atmospheric variations impact the \om-\oz\ relationship. Using such a large range of \no\ allows us to understand in which scenarios we would expect \oz\ abundances to be impacted in a way that would make observations more difficult to interpret. We briefly review expected \no\ surface fluxes over time, and  atmospheric parameters impact their mixing ratios.

As alluded to earlier, abundances of \no\ can be strongly impacted by \om\ abundances due primarily to UV shielding and reactions with \om\ and oxygen-containing species. In addition, the level to which \no\ can buildup in an atmosphere is also dependent on the host star, with cooler stars allowing for more buildup in their atmospheres due to lower incoming UV fluxes  (e.g., \citealt{segu03,segu05}). As mentioned during the previous subsection, with \no\ there has been work done to evaluate the maximum possible surface fluxes and corresponding mixing ratios using biogeochemical and photochemistry models \citep{schw22}. It is also more complicated to evaluate these limits, since \om\ influences both the surface flux and mixing ratio of \no. The production of \no\ surface flux is caused by nitrification and denitrification processes, whereas \no\ mixing ratios depend on \no\ destruction rates via photolysis and \od\ radicals in the stratosphere. For both surface flux and atmospheric mixing ratios, the dependence on \om\ is complicated, as nitrification processes require \om\ and denitrification processes require an absence of \om. Moreover, while \om\ also creates \oz, which protects atmospheric \no\ from photolysis, increased \om\ also causes increased \od\ radicals -- another major sink for \no. Additionally, due to the dependency on incoming UV, the amount of \no\ in the atmosphere is highly influenced by the host star, requiring careful modeling.

It is possible that \no\ surface fluxes were much higher in the past, particularly during the Proterozoic era. On modern Earth the majority of denitrification processes end with \no\ being converted into N$_2$, but this conversion requires a metal catalyst, most commonly Copper (see \citealt{schw22} for details). However during the Proterozoic era it is estimated that the oceans were highly depleted of Copper \citep{sait03,zerk06,robe11}, potentially causing significantly higher \no\ fluxes as \no\ would not be converted into N$_2$. Although, for planets around hotter stars like the Sun in order not to have widespread depletion of \no\ via photolysis, \om\ levels of about 10\% PAL would be necessary to provide UV shielding \citep{robe11}. If so, \no\ could have accumulated on Proterozoic Earth to levels high enough to contribute to warming during this time period.

\section{Summary and conclusions \label{sec:conclusions}}
This study focuses on how varying \no\ abundances in the atmosphere of an Earth-like planet impact the \om-\oz\ relationship across a range of \om\ levels around a variety of host stars. We find that the impact of varying \no\ depends on both the host star and the amount of \om\ in the atmosphere (see Fig.~\ref{fig:O2O3_relationship_grids}). Adding additional \no\ to an atmosphere rather than removing it consistently yielded more significant changes to \oz\ formation and destruction.

Atmospheric chemistry for models with varying \no\ (Sect.~\ref{sec:N2O_chem}) show that, for \om\ levels greater than 1\% PAL, planets around all hosts except the M5V experience significant depletion in \oz\ in the high \no\ models compared to modern levels of \no, due to the  enhanced efficiency of the \nox\ catalytic cycle. Planets around the M5V hosts are the least impacted by this effect, due to their low-incoming UV flux having a weakened ability to convert \no\ into \nox. However, the M5V-hosted planets were most efficient at creating extra \oz\ at high \no\ with the smog mechanism in the lower atmosphere, especially at lower \om\ levels as the low UV environment was more suited to \noo\ photolysis than \om\ photolysis. At \om\ levels of 0.1\% PAL and lower, planets around all hosts experienced an increase in \oz\ with the high \no\ models. This was due to the more dominant effects of the smog mechanism, as the Chapman mechanism was extremely limited by low \om. However, the increase was not as pronounced as that seen for the planet around the M5V host. This is because the higher efficiency of hotter stars in converting \no\ into \nox\ resulted in \nox\ abundances high enough to reach the \nox-saturated regime, thereby suppressing \oz\ smog formation.

The UV flux reaching the surface of our model planets was impacted by the changes in \oz\ (Sect.~\ref{sec:UV_to_ground}) especially since \oz\ abundance and attenuation of UV in the atmosphere have a nonlinear relationship. Results for the UVC surface flux are shown in Fig.~\ref{fig:UVC_barchart}, as this wavelength range is not only the most dangerous for life, but also the most sensitive to changes in \oz. The most extreme example of changing UVC surface flux was observed for the G0V-hosted planet at modern levels of \om\ and high \no, which received a 15 billion-fold increase in UVC ground flux compared to modern levels of \no. For planets around all the hosts except the M5V, the largest changes in UVC at the surface occurred in the high \no\ cases. More flux reached the surface at \om\ levels above 0.1\% PAL due to \nox\ destroying \oz, while less UV reached the ground in low \om\ cases of 0.1\% PAL because of additional UV shielding from \oz\ produced by the smog mechanism. However, for the high \no\ models around the M5V host, there was greater UV shielding at \om\ levels of 10\% PAL and below, due to the increased \oz\ formed by the smog mechanism.

Lastly, we explored how changing \no\ abundances could potentially impact \oz\ measurements in future observations and the ability to use \oz\ to learn about the amount of \om\ in a planetary atmosphere (Sect.~\ref{sec:emission_spectra}). We focused particularly on the 9.6~$\mu$m \oz\ feature for the simulated planetary emission spectra and how the feature would change with different abundances of \no\ (Fig.~\ref{fig:emission}). Unlike in \cite{koza22}, changes in feature depth were influenced more by changes in \oz\ than by temperature profiles. While decreasing the amount of \no\ had little impact on the \oz\ feature, increasing it had a significantly larger impact. The destruction of \oz\ for hotter hosts at high \om\ levels particularly influenced the feature depth, resulting in shallower features, with the K5V-hosted planet being the most impacted. The planet around the M5V dwarf only had significant changes in the \oz\  feature for the high \no\ models at low \om\ levels, due to the amount of extra \oz\ produced by the smog mechanism in the lower atmosphere.

When considering this work in the context of planning future observations, no host star displayed results showing that  \oz\ measurements in the mid-IR would be unaffected by the abundance of \no\ in their atmospheres. We propose that a separate measurement of \no\ is important for using \oz\ to assess the abundance of \om\ in an atmosphere. Fortunately, if we consider mid-IR measurements once again, there exists a \no\ feature, albeit overlapping with a \ch\ feature, potentially causing data interpretation issues. A variety of studies have been carried out in preparation for future observations of the proposed LIFE mission in the mid-IR (e.g., \citealt{konr22,alei22,konr23,alei24,ange24}); however, more in-depth studies focusing on the different levels of \no\ would be helpful in the pursuit of using \oz\ as a means to learn about potential life on a terrestrial exoplanet.

Studying the \om-\oz\ relationship in the context of varying \no\ abundances reveals another layer of complexity in addition to what we already find in \cite{koza22}. As with all atmospheric measurements of terrestrial exoplanets, context will be essential for truly understanding the meaning of our observations. This work further reinforced the idea that understanding the host star is necessary to understand the planet. It also highlights the need to measure additional atmospheric species before using \oz\ as a way to infer biologically produced \om\ and the potential for surface life on a planet.

\begin{acknowledgements}
All computing was performed on the HPC cluster at the Technical University of Denmark \citep{hpc}.  This project is funded by VILLUM FONDEN. Authors TK and LML acknowledge financial support from the Severo Ochoa grant CEX2021-001131-S funded by MCIN/AEI/ 10.13039/501100011033. JMM acknowledges support from the Horizon Europe Guarantee Fund, grant EP/Z00330X/1. We thank the anonymous referee for providing comments, which improved the clarity of our manuscript.
\end{acknowledgements}

\bibliographystyle{aa}
\bibliography{main.bib}{}

\begin{appendix}
\onecolumn
\section{Supporting figures and tables \label{append:A}}

\begin{figure*}[h!]
\centering
\includegraphics[scale=0.55]{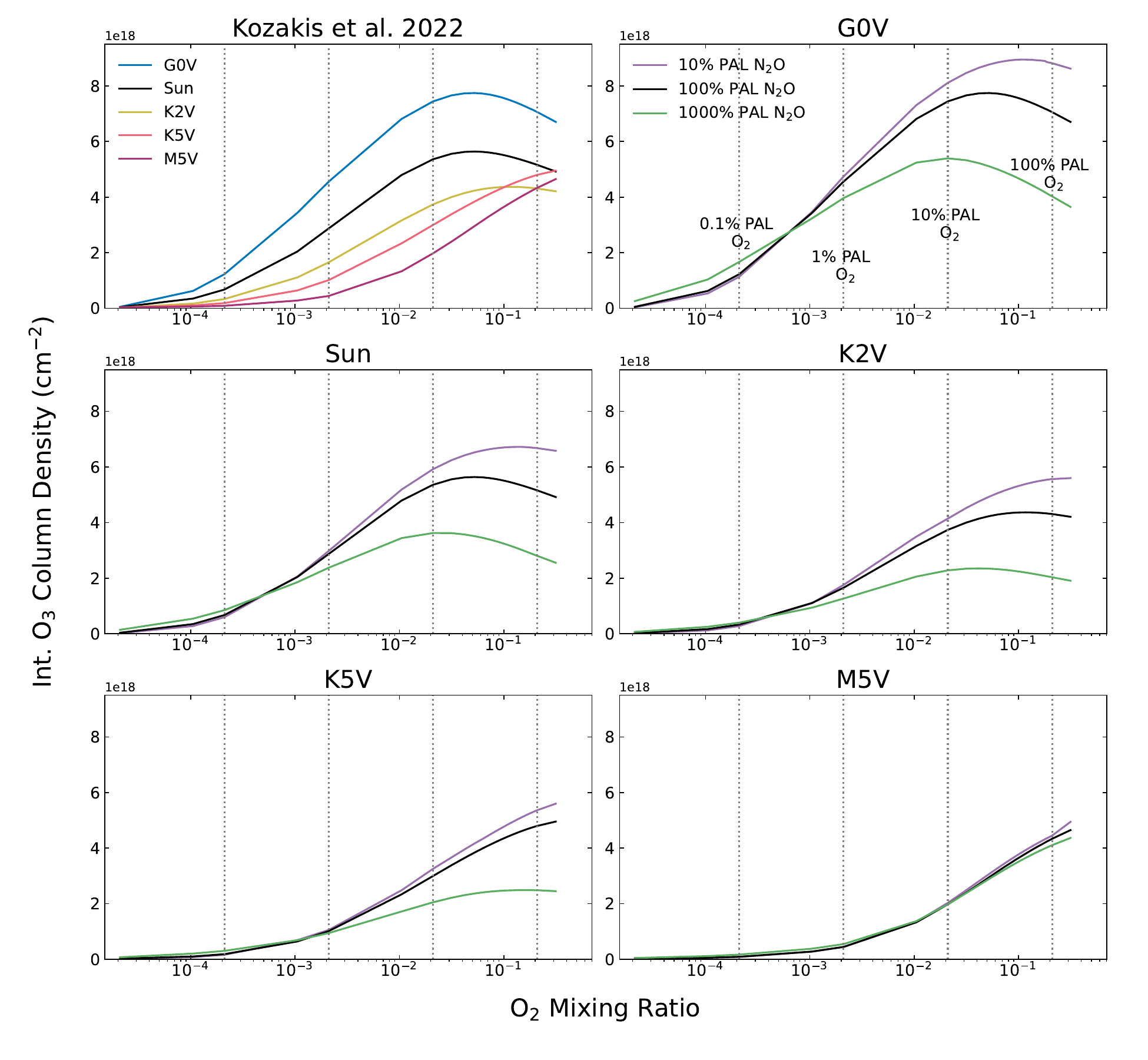}
\caption{\om-\oz\ relationship for changing \no, along with models from \cite{koza22} for comparison, for all modeled \om\ mixing ratios. Vertical dashed lines indicate \om\ levels of 100\%, 10\%, 1\%, and 0.1\% PAL. All plots share the same y-axis to facilitate comparison between different host stars. The phenomena that causes the peak \oz\ value for G0V-, Sun-, and K2V-hosted planets to occur at \om\ levels less than the maximum value modeled (150\% PAL \om) is discussed at length in \cite{koza22}.
\label{fig:O2O3_relationship}}
\end{figure*}

\begin{figure*}[h!]
\centering
\includegraphics[scale=0.35]{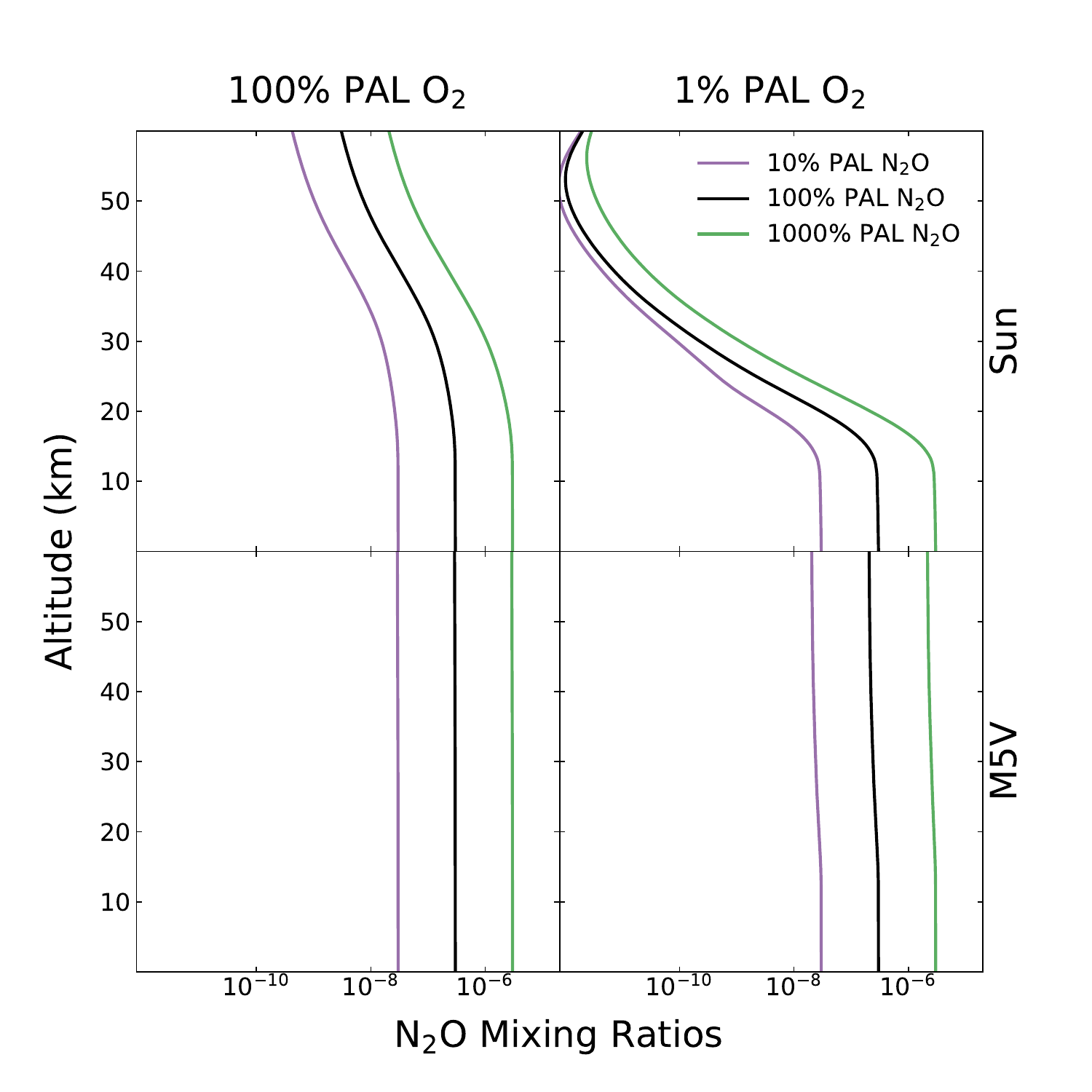}
\caption{\no\ profiles for planets around the Sun and M5V hosts at 100\% and 1\% PAL \om. The sun-hosted planet shows significant amounts of \no\ depletion via photolysis (the main stratospheric sink of \no), especially at lower \om\ levels, as the UV protection from both \om\ and \oz\ is significantly lessened. This depletion of \no\ impacts the conversion of \no\ to \nox. Meanwhile, the M5V-hosted planet experiences very little photolysis at all \om\ levels, due to the low UV flux of the host star.
\label{fig:N2O_photolysis}}
\end{figure*}

\begin{figure*}
\centering
\includegraphics[scale=0.4]{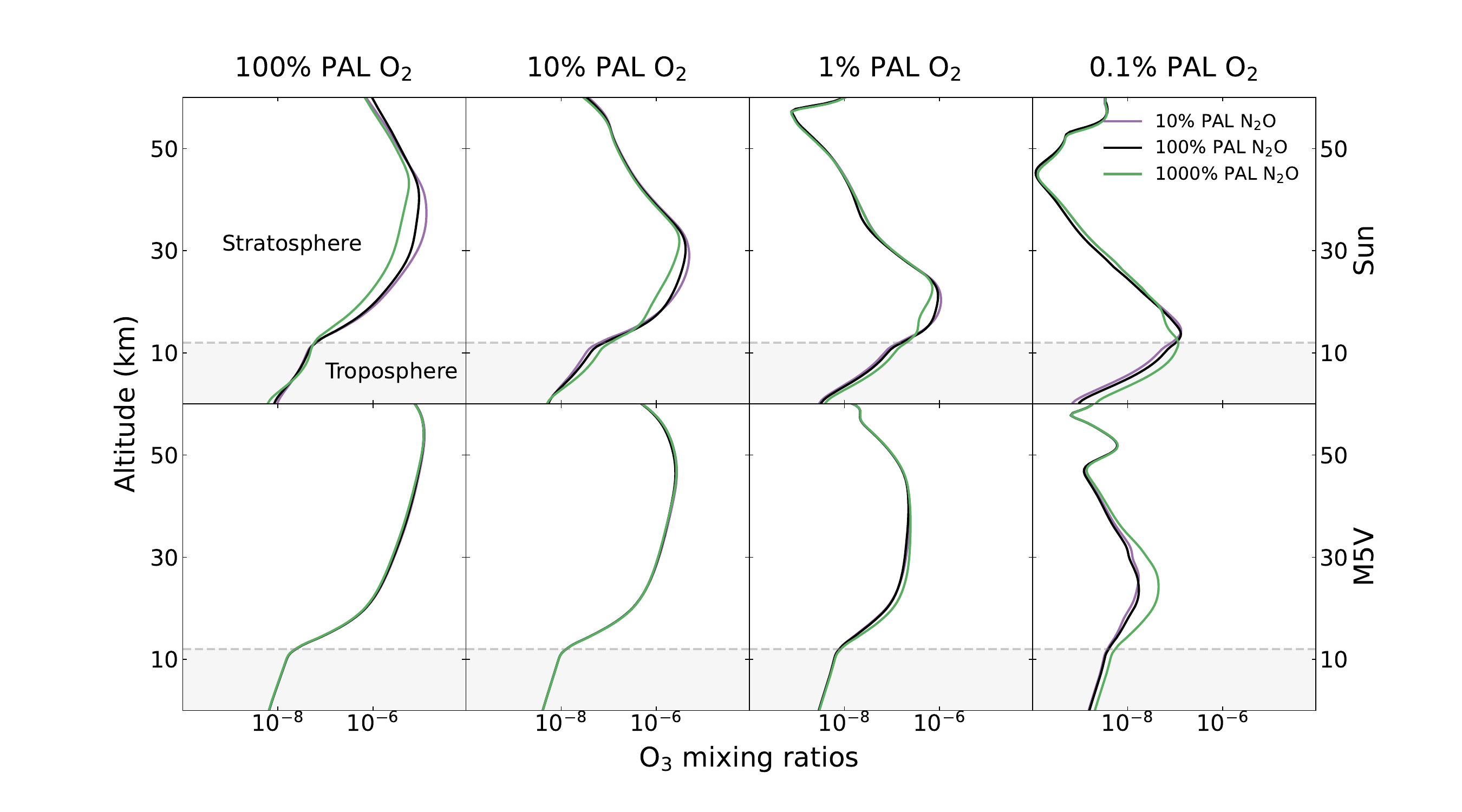}
\caption{\oz\ profiles for varying \no\ and \oz\ levels for planets around the Sun and the M5V hosts. The stratosphere (white background) and troposphere (gray background) are indicated in order to draw attention to differences in \oz\ production and depletion with the two stellar hosts. While at higher \om\ levels the Sun-hosted planet efficiently destroys stratospheric \oz, the M5V-hosted planet experiences minimal stratospheric \oz\ depletion at all \om\ levels. The smog mechanism dominates in the lower atmosphere for the two hosts, especially at lower \om, but it is restricted to the troposphere for the planet around the Sun. Smog \oz\ is produced into the stratosphere for the planet around the M5V host, due to the lack of suppression the Sun-hosted planet experiences in the \nox-saturated regime. See Figs.~\ref{fig:NOx_regimes} and~\ref{fig:NO_HO2} for more information on the \nox-limited and \nox-saturated regimes.
}
\label{fig:N2O_O3profiles}
\end{figure*}

{\singlespace
\begin{table*}[h!]
\centering
\footnotesize
\caption{UV Integrated Fluxes \label{tab:UV_all}}
\begin{tabular}{crrr|rr}
Spectral & O$_2$ MR & TOA flux & \cite{koza22}& \multicolumn{2}{c}{Surface Flux Normalized to \cite{koza22}}\\
\cline{5-6}
Type & (\% PAL) & (W/m$^2$) & Surface Flux (W/m$^2$)  & 10\% PAL \no\ & 1000\% PAL \no\  \\
\hline
\hline 
\multicolumn{6}{c}{UVB Fluxes (280 - 315 nm)}\\
\hline
G0V 	 & 100	 & 22.35	 & 1.5e+00	 & 0.72	 & 1.84	  \\
G0V 	 & 10	 & 22.35	 & 1.4e+00	 & 0.88	 & 1.47	 \\
G0V 	 & 1	 & 22.35	 & 2.4e+00	 & 0.97	 & 1.13	  \\
G0V 	 & 0.1	 & 22.35	 & 5.9e+00	 & 1.05	 & 0.84	  \\
\hline
Sun 	 & 100	 & 16.18	 & 1.6e+00	 & 0.74	 & 1.69	 \\
Sun 	 & 10	 & 16.18	 & 1.6e+00	 & 0.89	 & 1.43	  \\
Sun 	 & 1	 & 16.18	 & 2.7e+00	 & 0.98	 & 1.13	  \\
Sun 	 & 0.1	 & 16.18	 & 5.6e+00	 & 1.05	 & 0.90	 \\
\hline
K2V 	 & 100	 & 4.8	 & 6.8e-01	 & 0.79	 & 1.67	  \\
K2V 	 & 10	 & 4.8	 & 7.4e-01	 & 0.92	 & 1.39	  \\
K2V 	 & 1	 & 4.8	 & 1.2e+00	 & 0.97	 & 1.13	  \\
K2V 	 & 0.1	 & 4.8	 & 2.2e+00	 & 1.04	 & 0.95	  \\
\hline
K5V 	 & 100	 & 0.68	 & 9.8e-02	 & 0.90	 & 1.60	  \\
K5V 	 & 10	 & 0.68	 & 1.4e-01	 & 0.95	 & 1.24	  \\
K5V 	 & 1	 & 0.68	 & 2.3e-01	 & 0.99	 & 1.02	  \\
K5V 	 & 0.1	 & 0.68	 & 3.4e-01	 & 1.02	 & 0.92	  \\
\hline
M5V 	 & 100	 & 3.5e-02	 & 6.5e-03	 & 0.98	 & 1.04	  \\
M5V 	 & 10	 & 3.5e-02	 & 1.0e-02	 & 0.99	 & 1.00	  \\
M5V 	 & 1	 & 3.5e-02	 & 1.6e-02	 & 1.00	 & 0.96	  \\
M5V 	 & 0.1	 & 3.5e-02	 & 2.0e-02	 & 1.01	 & 0.94	  \\
\hline
\hline
\multicolumn{6}{c}{UVC Fluxes (121.6 - 280 nm)} \\
\hline
G0V 	 & 100	 & 11.2	 & 3.8e-27	 & 2.3e-02	 & 1.5e+10	  \\
G0V 	 & 10	 & 11.2	 & 1.8e-08	 & 5.9e-01	 & 3.1e+00	  \\
G0V 	 & 1	 & 11.2	 & 1.7e-04	 & 1.0e+00	 & 5.2e-01	  \\
G0V 	 & 0.1	 & 11.2	 & 1.7e-02	 & 1.3e+00	 & 1.9e-01	  \\
\hline
Sun 	 & 100	 & 6.7	 & 2.8e-21	 & 4.1e-06	 & 8.8e+08	  \\
Sun 	 & 10	 & 6.7	 & 3.3e-08	 & 6.0e-01	 & 3.9e+00	  \\
Sun 	 & 1	 & 6.7	 & 2.3e-04	 & 1.0e+00	 & 7.2e-01	  \\
Sun 	 & 0.1	 & 6.7	 & 1.5e-02	 & 1.4e+00	 & 3.2e-01	  \\
\hline
K2V 	 & 100	 & 1.4	 & 1.1e-18	 & 2.0e-05	 & 6.0e+08	  \\
K2V 	 & 10	 & 1.4	 & 1.8e-08	 & 6.8e-01	 & 4.1e+00	  \\
K2V 	 & 1	 & 1.4	 & 1.0e-04	 & 9.8e-01	 & 8.4e-01	  \\
K2V 	 & 0.1	 & 1.4	 & 1.0e-02	 & 1.7e+00	 & 4.6e-01	  \\
\hline
K5V 	 & 100	 & 0.16	 & 2.1e-21	 & 7.8e-03	 & 6.6e+08	  \\
K5V 	 & 10	 & 0.16	 & 8.4e-09	 & 7.8e-01	 & 2.3e+00	  \\
K5V 	 & 1	 & 0.16	 & 5.4e-05	 & 1.0e+00	 & 5.7e-01	  \\
K5V 	 & 0.1	 & 0.16	 & 4.7e-03	 & 1.3e+00	 & 2.8e-01	  \\
\hline
M5V 	 & 100	 & 2.7e-02	 & 9.8e-21	 & 3.7e-01	 & 7.4e+00	  \\
M5V 	 & 10	 & 2.7e-02	 & 4.1e-09	 & 9.7e-01	 & 5.9e-01	  \\
M5V 	 & 1	 & 2.7e-02	 & 3.8e-05	 & 1.0e+00	 & 4.0e-01	  \\
M5V 	 & 0.1	 & 2.7e-02	 & 1.1e-03	 & 1.1e+00	 & 3.6e-01	  \\
\hline
\hline
\end{tabular}
\vspace{-0.2cm}
\tablefoot{
Abbreviations: MR = mixing ratio; PAL = present atmospheric level; TOA =  top of atmosphere}\end{table*}
}

\end{appendix}

\end{document}